*Type of Paper (Review)*

# Health Digital Twins Supported by Artificial Intelligence-based Algorithms and Extended Reality in Cardiology

**Zofia Rudnicka[1], Klaudia Proniewska [2,3], Mark Perkins [4,5], and Agnieszka Pregowska [1,*]**

[1] Institute of Fundamental Technological Research, Polish Academy of Sciences, Pawinskiego 5B, 02-106, Warsaw, Poland; zrudnick@ippt.pan.pl, aprego@ippt.pan.pl
[2] Center for Digital Medicine and Robotics, Jagiellonian University Medical College, 7E Street, 31-034 Krakow, Poland; klaudia.proniewska@uj.edu.pl, peter.van.dam@peacs.nl
[3] Department of Bioinformatics and Telemedicine, Jagiellonian University Medical College, Medyczna 7 Street, 30-688 Krakow, Poland; klaudia.proniewska@uj.edu.pl
[4] Collegium Prometricum, the Business School for Healthcare, 136 81-701 Sopot, Poland; mark.perkins@prometricum.eu
[5] Royal Society of Arts, 8 John Adam St, London WC2N 6EZ, United Kingdom; mark.perkins@prometricum.eu

[*] Correspondence: aprego@ippt.pan.pl; Tel.: +48-22-826-1281 (ext. 412)

**Abstract:** Recently, significant efforts have been made to create Health Digital Twins (HDTs), digital twins for clinical applications. Heart modeling is one of the fastest-growing fields, which favors the effective application of HDTs. The clinical application of HDTs will be increasingly widespread in the future of healthcare services and has a huge potential to form part of the mainstream in medicine. However, it requires the development of both models and algorithms for the analysis of medical data, and advances in Artificial Intelligence (AI) based algorithms have already revolutionized image segmentation processes. Precise segmentation of lesions may contribute to an efficient diagnostics process and a more effective selection of targeted therapy. In this paper, a brief overview of recent achievements in HDT technologies in the field of cardiology, including interventional cardiology was conducted. HDTs were studied taking into account the application of Extended Reality (XR) and AI, as well as data security, technical risks, and ethics-related issues. Special emphasis was put on automatic segmentation issues. It appears that improvements in data processing will focus on automatic segmentation of medical imaging in addition to three-dimensional (3D) pictures to reconstruct the anatomy of the heart and torso that can be displayed in XR-based devices. This will contribute to the development of effective heart diagnostics. The combination of AI, XR, and an HDT-based solution will help to avoid technical errors and serve as a universal methodology in the development of personalized cardiology. Additionally, we describe potential applications, limitations, and further research directions.

**Keywords:** Artificial Intelligence; Machine Learning; Metaverse; Virtual Reality; Extended Reality; Augmented Reality; Digital Twin; Health Digital Twin; personalized medicine; cardiology;

## 1. Introduction

A Digital Twin (DT) is a digital replica of its corresponding physical object or process. It is a virtual model with special features that combine the physical and digital worlds [1]. Since modern medicine needs to move from being a wait-and-response therapeutic discipline to an interdisciplinary preventive science, interest in the application of DT technology in medicine is rapidly growing [2]. DTs enable human physical characteristics, including changes and disorders in the body, to be transferred to the digital environment. Thus, DT technology also opens up the possibility of delivering personalized medicine in the form of providing an individual patient with their very own diagnosis, optimization path, health forecast, and treatment plan [3]. Thus, Health Digital Twins (HDTs) may represent a specific organ modeled from high-resolution medical imaging and structural and physiological functional data across multiple scales [4]. The technology can be applied to the development of drug delivery processes, selection of targeted therapies, and design of clinical trials. HDTs fit perfectly into the Healthcare 4.0 concept that assumes the introduction of a publicly available system of effective personalized health care [5].

One of the technologies that is applied in the implementation of DTs is Extended Reality (XR) [6]. This enables users to experience the feeling of immersion in the real world on various levels through head-mounted displays (HMDs) [7]. This approach provides a new level of quality in the three-dimensional (3D) visualization of complex structures such as organs and their abnormalities as well as touch-free interfaces [8]. XR is increasingly used in preoperative planning, and recently even during surgery [9,10]. Immersive solutions are also starting to play an important role in medical education [11], particularly in the context of distance education [12]. Thus, in the case of a virtual environment and virtual models, the key issue is the development of an environment and/or scene and a model that reflects reality as closely as possible. In this field, Artificial Intelligence (AI)-based algorithms, in particular, Deep Neural Networks (DNNs), have recently revolutionized image creation [13]. Precise segmentation of lesions may contribute to an efficient diagnostics process and a more effective selection of targeted therapy. For example, an AI-based algorithm for the segmentation of pigmented skin lesions has been developed, which enables diagnosis in the earlier stages of the disease, without invasive medical procedures [14,15]. With flexibility and scalability, AI can be also considered an efficient tool for cancer diagnosis, particularly in the early stages of the disease [15,1617]. On the other hand, in this context, the provision of a stable internet connection is extremely important. However, different XR-based solutions have varying requirements for optimal connectivity. Thus, intelligent DTs combined with AI-based algorithms and XR devices have a huge potential to revolutionize medicine and public health. A basic model of such connections is presented in **Figure 1**.

In this paper, we present a brief overview of the recent achievements in Health Digital Twin technologies in the field of cardiology, taking into account the application of Extended Reality and Artificial Intelligence. Specifically, we aim to answer the following research questions (RQs).

- ***RQ1:*** *Can AI-based algorithms be used for the accurate segmentation of human organs based on medical data, in particular in the case of the heart?*
- ***RQ2****: How do AI-based algorithms be beneficial in the Health Digital Twin technologies?*
- ***RQ3****: How can Extended Reality be used in Health Digital Twin-based solutions?*
- ***RQ4****: What ethical threats does a world based on the Metaverse and Artificial Intelligence pose to us?*

To address the above research questions, a systematic literature review was conducted. The paper is structured as follows. **Section 1** comprises the introduction. **Section 2** describes the research methodology. **Section 3** presents the application of DTs in the field of medicine with a special emphasis on cardiology. The potential of Extended Reality in medicine is shown in **Section 4**, while **Section 5** focuses on the application of AI in the context of the Metaverse and Digital Twins. **Section 6** presents data security issues. Ethical issues and risks are described in **Section 7**. Finally, **Section 8** presents a discussion and the main conclusions of the study.

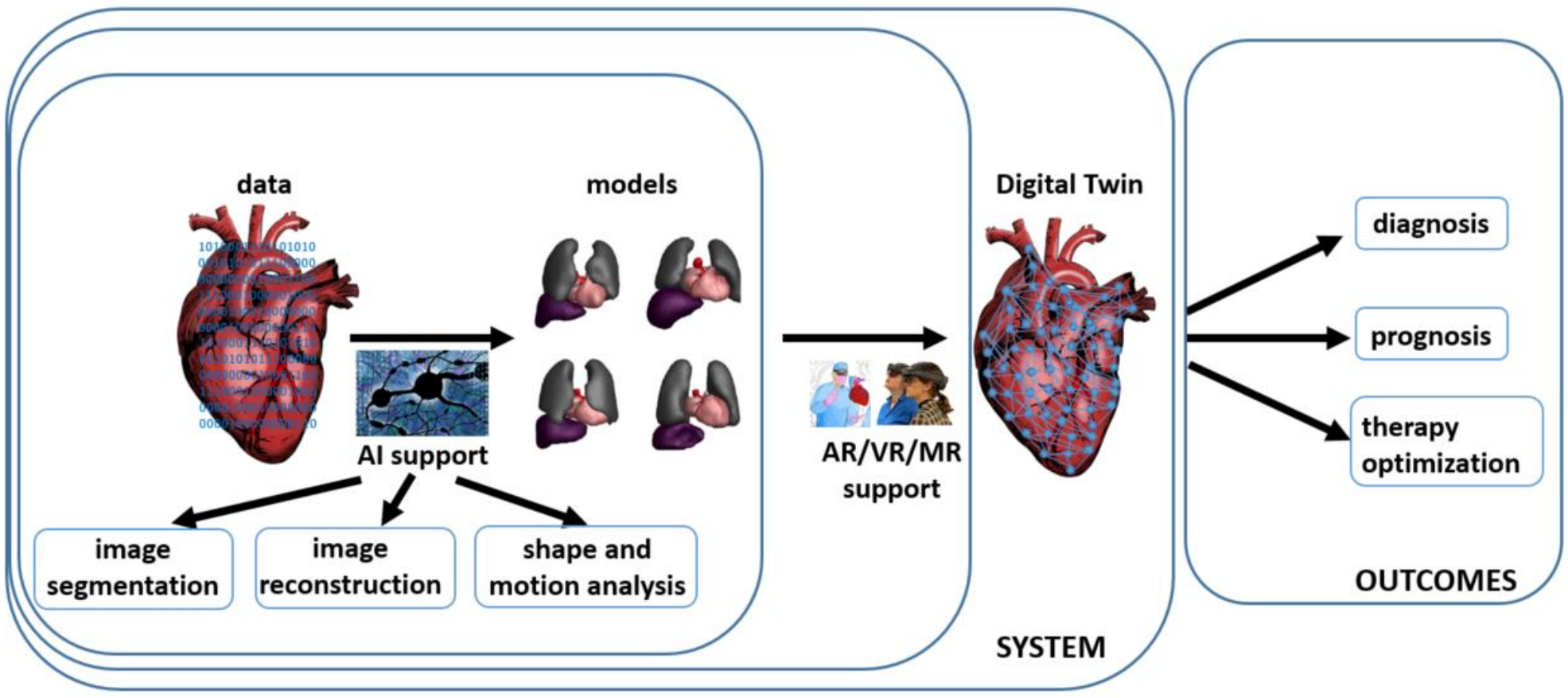

**Figure 1**. Health Digital Twin supported by Artificial intelligence and Extended Reality: a basic workflow.

## 2. Materials and Methods

In this paper, a systematic review was undertaken based on the PRISMA Statement which has been published in several journals, and its extensions, including PRISMA-S [18] formed to accommodate the wide question concerning the application of Digital Twin technology supported by Artificial intelligence and Extended Reality-based solutions in the field of cardiology. Eligible materials, including publications, reports, protocols, and papers from peer-reviewed literature were identified from Scopus, Web of Science, and PubMed databases. The keywords: artificial intelligence, machine learning, digital twin, digital twin in medicine, digital twin in cardiology, extended reality, mixed reality, virtual reality, augmented reality, metaverse, digital heart, cardiology, signal segmentation, medical image scan segmentation, segmentation algorithms, classification algorithms, and their variations were used. The inclusion criteria set to select the resources were as follows:

- *resource language*: English,
- *type of resource*: publications in the form of journal papers, books, and proceedings as well as technical reports

and

- *a publishing time frame between 2020 and 2024*. This choice of publication dates is based on the fact that areas such as artificial intelligence, the metaverse, and health digital twins have developed very dynamically in recent years, and we wanted to focus only on the latest developments in the field of the metaverse and digital twins taking into account a medical context, in particular cardiology.

The selected sources were analyzed first in terms of compliance with the topic analyzed, and then their contribution to it. In the first step, the material was evaluated taking into account the title and abstract. Duplicated records were removed from consideration.

To the study, the following exclusion criteria were applied.

- *Resources such as PhD theses were omitted*
- *Material not related to artificial intelligence, the Metaverse, and HDTs in the field of medicine, in particular cardiology, were removed*.

In addition, resources that were deemed unnecessary during the search were eliminated from consideration. Thus, all documents considered needed to be peer-reviewed and to include answers to research questions. Finally, 335 documents were taken into account.

## 3. Digital Twin Technology in Medicine: Health Digital Twins

A digital twin contains three parts: the physical model of the organ/body, their virtual counterpart, and their mutual interactions. For example, this approach works well in industrial solutions such as the battery industry [19]. This allows for a pro-ecological approach to batteries already at the design stage. In turn, in the area of medicine, DTs offer opportunities ranging from research on mechanisms related to various diseases and isolating disease predictors to optimization of health outcomes [20]. However, the human body and its parts are more complex than objects in engineering and manufacturing [21, 22]. It has also been shown that mathematical models can be calibrated based on patient-specific data to predict tumor response dynamics [23,24,25,26]. These models can be applied in the development of a digital twin of the patient. In recent years, early attempts to apply DTs in medicine have been made [27]. Cho et al. [28] have applied digital twins to endodontic treatment and they have been applied to cold treatment by Michaux et al. [29]. DTs have been also applied to cancer treatment [3031]. In cancer cases, the digital twin makes it possible to analyze the patient's response to the simulated development of tumors under different treatments, which is not possible in the real world [32]. As a consequence, the physician has the opportunity to select the most efficient cancer therapy path: this is personalized in that the DT is based on a biological model applied to a specific patient's data. Another interesting application of DT technology has been demonstrated by Croatti et al [33], namely trauma management. A virtual world is presented, that of the traumatized patient and his surroundings. This enables the inclusion of contextual information in the treatment process, such as the impact of the medical staff's approach to the patient. Thus, while DT solutions may become a clinical reality, virtual models of real organizations are also needed. Moreover, HDTs in medicine are at an early stage of development and must be properly verified before they can enter clinical practice.

### *3.1 Digital Twins in Cardiology – the Heart Digital Twin*

In cardiology, DTs solve the inverse problem of electrocardiography, relating electrical signals to the anatomy of the heart [33,34], connected with the development of a heart model that involves the parameterization of its elements [35]. This enables non-invasive functional cardiac modeling imaging methods to be used in a clinical setting. For this to become a clinical reality 3D thorax information should result in well-defined heart models inside the thorax, for which currently no algorithms are available due to the large variability in body build and underlying heart disease. However, the solution to the inverse problem is subject to technical errors [36]. For further development of this DT, a model of the patient's heart and torso derived from medical imaging based on Magnetic Resonance Imaging/Computer Tomography (MRI/CT) is required [37]. Medical image processing is time-consuming, especially when the image is of poor quality. Moreover, the participation of humans is needed. Thus, the development of efficient and accurate segmentation algorithms is of high importance [38].

**Table 1.** Comparison of recent (2020-2024) developments in the application of XR in cardiology and interventional cardiology.

| XR technology type | Head-mounted display (HDM) type | AI support | Perception of real surrounding | Application Field | References |
|---|---|---|---|---|---|
| MR | HoloLens 2 | No | Yes | Visualization of ultrasound-guided femoral arterial cannulations | 39 |
| MR | HoloLens 2 | No | Yes | USG visualization | 40 |
| MR | HoloLens 2 | No | Yes | Visualization of heart structures | 41 |
| MR | HoloLens 2 | No | Yes | Operation planning | 42 |
| MR | HoloLens | No | Yes | Operation planning | 43 |
| MR | HoloLens 2 | No | Yes | Visualization of heart structures | 44 |
| MR | HoloLens 2 | No | Yes | Visualization of heart structures | 45 |
| AR | mobile phone | No | Yes | Diagnosis of the heart | 46 |
| AR | none | No | Yes | Virtual pathology stethoscope detection | 47 |
| AR | none | Yes | Yes | Eye tracking system | 48 |
| AR | none | Yes | Yes | Detection of semi-opaque markers in fluoroscopy | 49 |
| VR | Simulator Stanford Virtual Heart | No | No | Visualization of heart structures | 50 |
| VR | Simulator Stanford Virtual Heart | No | No | Visualization of heart structures | 51 |
| VR | Meta-CathLab (concept) | No | No | Merging interventional cardiology with the metaverse | 52 |
| VR | VR glasses | Yes | No | Sleep stage classification - concept | 53 |
| VR | Virtualcpr: mobile application | Yes | No | Training in cardiopulmonary resuscitation techniques | 54 |
| VR | none | Yes | No | Diagnostic of cardiovascular diseases - visualization | 55 |
| VR | none | No | No | Cardiovascular Education | 53 |

## 4. **Extended Reality in Medicine**

Amongst other things, XR in medicine provides the ability to overlay computer-developed elements and structures onto real-world data and develop touch-free human-computer interfaces (controlled by voice, eye movements, and hand gestures) that can be applied in a sterile environment [56]. Moreover, Schöne et al. [57] have shown that the current configuration of XR-based devices contributes to the experience of a feeling of reality, although this depends on the quality of the scenes and objects presented. Extended reality can be divided into Virtual Reality (VR) (a completely virtual experience), Augmented Reality (AR) (combining real-world elements with computer-generated elements), and Mixed Reality (MR) (computer-generated elements that can actively interact with the real world) [4]. All these types of XR have been provisionally applied in operation planning [58,59], rehabilitation [60], and medical education [61,62]. The potential of XR technologies in anxiety reduction has been demonstrated by Yan et al. [63] and Ribé-Viñes et al. [64]. Additionally, Evans et al. [65] have proposed an evaluation of the physiological fidelity of the MR-based trauma situation. This may contribute to the development of more realistic solutions. Surprisingly, it has been shown that VR can reduce pain, in particular in the case of women in labor [66]. ER can also be combined with rehabilitation robots to increase patient motivation [67,68]. AR-based solutions are mostly limited to the monitoring of health status [68,69,70] and medical education [71]. VR predominates in medical simulators [72] and MR in medical education [73], especially in the study of anatomy, as well as in the planning of surgical procedures. The latter activity in particular allows for navigation during surgery [74] and observance of the course of the operation without disturbing the operating surgeon [75]. Furthemore, the operating surgeon may also consult another doctor located in a different location [76]. It also allows both medical students and medical staff to improve their skills. Furthermore, an interesting application of MR-based technologies has been proposed by Black et al. [77], namely human tracking for tele ultrasound.

### *4.1 Extended Reality in Cardiology*

Recently, one has observed exponential growth in cardiac imaging technologies [78, 79]. Congenital heart disease (CHD) has been shown to have a very complex three-dimensional anatomy. As a consequence, it is difficult to diagnose. This is where XR-based technology can be very helpful [80,81,82] (also see **Table 1**). Marvin et al. [83] have shown that this approach is quite popular in European cardiology, although it has not led to routine practice. For example, VMersive (VR-Learning, Poland) is an automated tool dedicated to the reconstruction of CT and MRI image scenes. It can be applied to procedure planning as in congenital heart disease treatment [84]. Another study [85] concentrates on the evaluation of a VR-based solution for baffle planning in CHD. The proposed approach enables a medical doctor to simulate different baffle configurations and analyze their impact on blood flow, which is under practical conditions impossible. The knowledge gained may be beneficial in operational planning [86]. In turn, Ghosh et al. [87] applied VR to the creation of a 3D view based on cardiac MRI to visualize multiple ventricular septal defects, an approach that uses commercial software to segment heart areas. This procedure made it possible to reveal what would normally be invisible, the ventricular septal defect of the heart. Another case of XR's practical application in cardiology is cardiac catheterization which is a commonly used procedure, although it is quite risky. Battal et al. [88] proposed VR as a support for this procedure. As shown by Eves et al. [89] and Chahine et al. [90] techniques such as AR may shorten operation time. When we consider cardiac surgery, even the simplest types, we must take into account the patient's recovery after such an operation. Very little attention is paid to this issue [91]. This creates a very interesting area of potential XR-based technology application [92]. Additionally, three-dimensional cardiac imaging is also important in veterinary medicine [93].

Another XR application field in cardiology is connected with rehabilitation. Mocan et al. [94] proposed a combination of home cardiac tele rehabilitation based on a virtual environment with a monitoring system. The idea presented allows the continuation of rehabilitation at home. AR may also be helpful in home monitoring as in an application that allows the correct placement of ECG electrodes to be checked using photos taken with a mobile phone [95]. Here it resulted that an AR-based solution enables at least eighty percent of the measurements to be obtained correctly.

Recently, attention has also been paid to pain management for surgical procedures as in cases of advanced heart failure where VR can also be beneficial [96]. Another approach to VR in cardiology is to use the sensors and cameras found in HDMs to determine Heart Rate (HR) [97]. HR was estimated based on the central portion of the face by application of remote photo plethysmograph, Eulerian Video Magnification (EVM), and Convolutional Neural Networks (CNNs). According to research results, it is only possible to predict HR based on facial regions.

A further important area of XR-based technology consists of the education (broadly understood) of both future medical staff and patients [98,99,100], as in the case of teaching heart anatomy [83]. In research by Patel et al. [101], traditional learning using 2D imaging and learning using XR and 3D imaging were compared. The objective of the study was to understand congenital heart disease. Although there were no differences in teaching effectiveness from a statistical point of view, participants who used XR in the learning process reported a better understanding of the content. On top of that, Lim et al. [102] found VR to be a very helpful tool for residents participating in pediatric cardiology rotations. O'Sullivan et al. [103] also found that more than eighty percent of participants believed that VR is a good teaching tool for acquiring knowledge about echocardiography, and over sixty percent of them rated VR higher than traditional teaching methods. Then, Choi et al. [104] found that AR glass increases the level of understanding of left ventricular ejection fraction. Additionally, García Fierros et al. [54] made a comparison of VR and fluoroscopic guidance for trans septal puncture. It turned out that VR may have the potential to shorten training. Gladding et al. [105] and Kieu et al. [50] also found that such an approach is helpful.

## 5. Artificial intelligence-Based Support in Medicine

Processing and analysis of biomedical data for diagnostic purposes is a multidisciplinary field that combines artificial intelligence, machine learning, biostatistics, time series analysis as well as statistical physics and algebra (including graph theory). Variables derived from biomedical phenomena can be described in several ways and in different domains (time, frequency, spectral values, and state spaces describing the biological system), depending on the characteristics and type of signal. Currently, due to the ability to process huge amounts of data, AI is beginning to play a key role in the processing of medical data [106]. The main AI application fields in medicine are connected with diagnosis processes, personalized treatment, patient care, and medical education [107]. Indeed, AI provides a huge support in decision-making processes. However, its use is not without technical and ethical challenges that must be solved for AI to become a significant part of clinical practice [108,109,110].

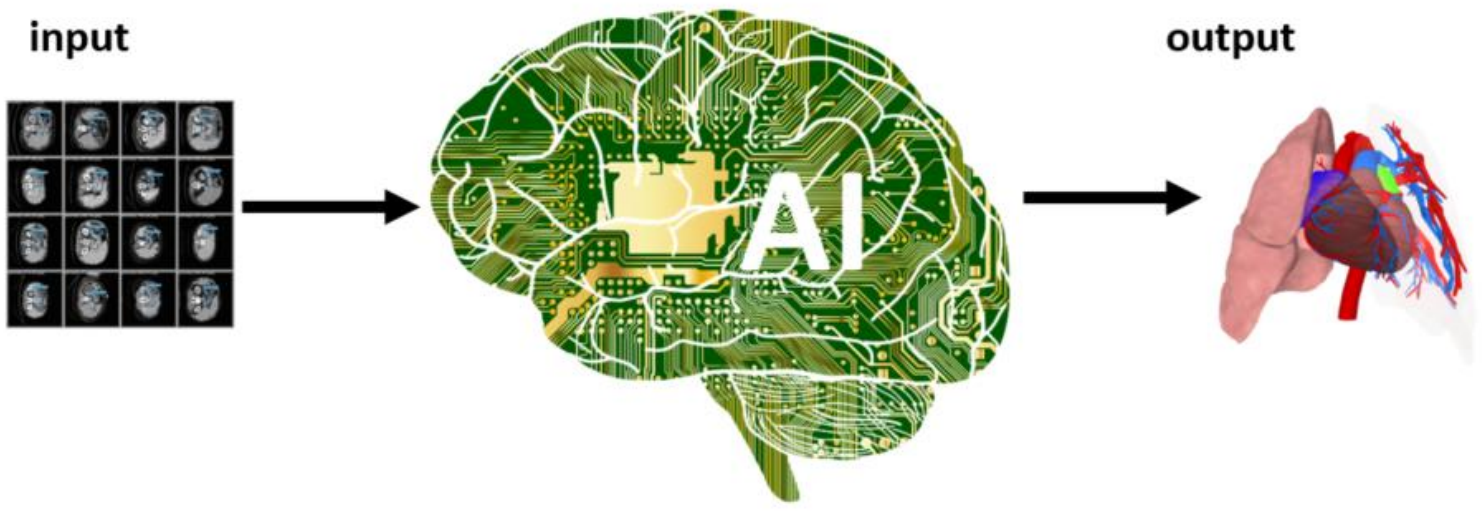

**Figure 2**. AI application in medical image segmentation: schema

### *5.1 Artificial Intelligence in Medical Image Segmentation*

Medical image segmentation is a significant area of AI application in medicine [111]. An image can be segmented in several ways, including semantic segmentation (the assignment of each pixel or voxel of an image to one of the classes) [112], instance segmentation (pixels of an image are assigned to the instances of the object) [113], and panoptic segmentation (the connection of the semantic and instance segmentation) [114]. The main disadvantage of semantic segmentation is the poor definition of the problem (sometimes multiple instances can be abstracted into a single class), which translates into inadequate recognition of image details. In the case of medical images, segmentation is often performed manually. This makes it a time-consuming and operator error-based process. Many algorithms have been proposed to support the automatic segmentation of medical images. The ultimate goal is to enable the fully automatic segmentation of any clinically acquired CT or MRI. For medical image segmentation (mostly semantic segmentation), the different types of neural networks, including Artificial Neural Networks (ANNs), Convolutional Neural Networks (CNNs), Recurrent Neural Networks (RNNs), Spiking Neural Networks (SNNs), Generative Adversarial Networks (GANs), Graph Neural Networks (GNNs), Sinusoidal Representation Networks (SIRENs) and Transformers, are applied [115].

#### *5.1.1 Application of the You-Only-Look-Once Algorithm*

The You-Only-Look-Once (YOLO) algorithm is an approach that is based on deep learning for object detection [116]. It depends on the idea that the images pass only once through the neural network, and hence the name. This is done by dividing the input image into a grid and predicting for each grid cell the bounding box and the probability of that class. The algorithm predicts different values related to the object, such as the coordinates of the center of the bounding box around the object, the height and width of the bounding box, the class of the object, and the probability, or the confidence of the prediction. This way of working may cause the algorithm to detect the object multiple times. To avoid duplicate detections of the same object the algorithm uses non-maximum suppression (NMS), which works by calculating a metric called Intersection over Union or (IOU) between the boxes. If the IOU between two boxes is larger than a certain threshold, the box with a higher confidence score is chosen and the other box is ignored. There have been many improvements to the YOLO algorithm. YOLO version 7 (YOLOv7), which was released in 2022, is currently in use [117]. It has several structural modifications, that provide higher accuracy, faster performance, improved scalability, and greater flexibility for customization. YOLOv7 operates in pixel space (2D). There is a need to extract the three-dimensional ROI for the pancreas (the dimensions of the ROI in the different anatomical planes (Axial, Coronal, Sagittal)), and combine this information to produce the 3D ROI.

In recent developments of the YOLO algorithm, there have been enhancements aimed at improving its efficiency and accuracy in medical imaging. These include the integration of advanced deep-learning techniques that enhance the algorithm's ability to discern finer details in complex medical images. For instance, the incorporation of convolutional neural network layers that are specifically tailored for medical image characteristics has led to more precise segmentation results [118]. Furthermore, adaptive thresholding techniques have been introduced to improve the performance of the IOU metric, making the algorithm more adept at handling varying shapes and sizes of medical structures. Another significant advancement is the integration of a 3D analysis capability. Unlike the traditional 2D grid approach, this new iteration can process three-dimensional data, which is crucial in accurately interpreting CT and MRI scans. This allows for a more detailed and comprehensive understanding of the scanned structures, leading to better diagnostics and treatment planning. Moreover, the latest versions of YOLO for medical imaging are being designed with an increased focus on user-friendliness and integration into existing medical workflows. This involves creating interfaces that are intuitive for medical professionals and ensuring compatibility with standard medical image formats and diagnostic tools. Such improvements aim to facilitate the wider adoption of this technology in clinical settings, ultimately contributing to more efficient and accurate patient care [119].

#### *5.1.2 Application of the Segment Anything Model Algorithm*

The most recent algorithm for image segmentation is the Segment Anything Model (SAM) which is based on a vision transformer (ViT) [120]. This solution allows pixel-level semantics and the position of target objects through hints to be indicated (points or boxes). This is a model pre-trained on big datasets. However, in the case of medical images, this training took place on a small data set [121,122,123]. In the case of medical image segmentation, SAM was able to find the boundaries of organs on CT/MRI image scans successfully but had difficulties in segmenting their anomalies [122]. For example, Zhou et al. [124] demonstrated that SAM can successfully segment polyps, but the segmentation was not always correct. In response, Liu et al. [125] added open-source 3D Slicer software to SAM to help catch errors. It is still the case, however, that in the field of medical image segmentation, SAM suffers from a low level of accuracy [121,126].

#### *5.1.3 Genetic Algorithms*

The analysis of medical data can also be approached using metaheuristic methods such as Genetic Algorithms (GAs), in particular, Evolutionary Algorithms (EAs) and Artificial Immune Systems (AISs) that search the possible solution space based on mechanisms taken from the theory of evolution and natural immune systems [127]. This approach can be applied to the analysis of the medical images that have a modality. EAs were successfully applied by Gudur et al. [128] to the classification of lung disorders based on chest X-ray images. It was used in the optimization of the parameters that then served as input to the CNNs. Arif & Wang [129] proposed a combination of GAs and the curvelet transform. These methods were applied to multimodal brain image analysis to enable more detailed information to be attached. This may contribute to more detailed medical image analysis. Moreover, EAs are used to encrypt medical data stored on servers to ensure their security [130], and another application field of EA concerns optimization of the drug delivery system [131].

#### *5.1.4 Artificial Neural Networks*

Artificial Neural Networks (ANNs) are networks whose structure and principle of operation are to some extent modeled on the functioning of fragments of the real nervous system (the brain) [132,133]. This computational invention contributes to the development of medical imaging, especially in cardiology, where their design, inspired by the human brain, enables them to interpret complex patterns within medical data effectively. ANNs consist of layers composed of a number of neurons, which apply specific weights and biases to the inputs. These neurons utilize nonlinear activation functions that enable the network to detect complex patterns and relationships that the linear functions might overlook. The output layer plays a pivotal role in making predictions or classifications based on the analysis, such as identifying signs of heart disease, classifying different cardiac conditions, or determining the severity of a disorder [134,135,136]. The properties of ANNs enable them to successfully process text, images, and videos [137]. For example, Binjun He et al. [138] focused on ANN-based lung cancer diagnosis. The diagnostics process began with identifying the lung cancer lesion area followed by the application of an image segmentation algorithm to display this area clearly. The AI model demonstrated a high level of accuracy in identifying lung cancer, showing powerful capabilities in both sensitivity and specificity.

#### *5.1.5 Convolutional Neural Networks*

Another neural network that has been applied to medical image processing is the Convolutional Neural Network (CNN). Opposite to traditional neural networks such as ANNs which typically process data in a straightforward, sequential manner, CNNs can discern spatial relationships within data sets. This is due to the way they were designed and constructed as they were intended to maintain and interpret the spatial structure of input data, an attribute that is vital for the accurate assessment of medical images. In other words, CNNs are composed of multiple layers, each designated for a specific task: convolutional layers, pooling layers, and fully connected layers. The crucial layer is the convolutional one that uses filters to methodically scan across the input image, capturing spatial features such as edges, corners, and textures. This is especially important in cardiology [139]. It enables the identification of intricate details in cardiac images, as it allows the network to progressively construct a more detailed representation of the image. This operation also enables the detection of features at different levels of abstraction, ranging from simple to complex structures. Pooling layers help in capturing the most essential features from the images, making the network's detection of features more robust, which is a crucial aspect in medical imaging where consistent feature identification across various cardiac images is vital. In turn, the fully connected layers take the high-level features filtered by the previous layers, flatten them out, and transform them into a final output. In the context of cardiac imaging, these layers are responsible for synthesizing the extracted features to identify specific conditions or anomalies in the heart. The final layer in a CNN is the output layer. For medical imaging tasks, this layer often uses a softmax activation function for classification tasks, such as when differentiating between types of heart diseases, or linear functions for regression tasks when estimating the severity of a condition [140]. Thus, these layers aid in making the network's detection of features more robust and sensitive to variations and distortions in the input image, a common challenge in cardiac images [141]. Therefore, fully connected layers play a crucial role in translating detailed features into actionable medical insights [142]. Another interesting application has also been introduced by Oyelade et al. [143]: the dual/twin Convolutional Neural Network (TwinCNN) for the analysis of mammogram images.

#### *5.1.6 Recurrent Neural Networks*

Recurrent Neural Networks (RNNs) are known for their ability to model long-term dependencies and are crucial for capturing the intricate details of cardiac structures. Unlike traditional feedforward neural networks that process inputs in a one-directional manner, RNNs are designed to handle sequences of data. This is achieved through their internal memory, which allows them to retain information from previous inputs and use it in the processing of new data [144]. The core function of RNNs lies in their ability to form a directed graph along a temporal sequence. This structure enables them to exhibit dynamic temporal behavior, essential for applications where time is a significant factor, such as in cardiac imaging. Each node in an RNN is capable of passing information to successive other nodes, creating a loop of information flow. This loop allows the network to remember previous information, an attribute not found in traditional neural networks. RNNs process inputs in a step-by-step fashion. At each step, the network not only considers the current input but also the information it has retained from previous inputs. This continuous integration of new and past data enables RNNs to make more informed decisions about the data being processed. For instance, in cardiac imaging, this means an RNN can analyze a sequence of heart images, taking into account the changes and patterns observed over time, rather than viewing each image in isolation. However, standard RNNs can struggle with long-term dependencies due to issues such as the vanishing gradient problem, where the influence of information decreases as it moves through the network [145,146]. Advanced variations of RNNs, such as Long Short-Term Memory (LSTM) and Gated Recurrent Units (GRUs), are crucial for complex tasks. LSTMs feature a unique gating system comprising input, forget, and output gates, allowing them to selectively retain or discard information, critical for managing the variable relevance of data in cardiac imaging. GRUs, on the other hand, simplify the LSTM structure by merging the input and forget gates into a single update gate, combining the cell state and hidden state for greater efficiency. This simplification results in faster training without compromising performance, making GRUs suitable for applications where both accuracy and computational efficiency are important [147].

#### *5.1.7 Spiking Neural Networks*

Most neural network models used in medicine use the simplest neuron model, namely the perceptron, a model containing simplified brain dynamics. The development of Spiking Neural Networks (SNNs) is intended to handle more biologically complex neuron models, so-called spiking neurons [148,149]. These neurons mimic how information is transferred in biological neurons including the precise time of occurrence of electrical impulses and so-called spikes or their sequences [150]. This approach is an alternative to traditional neural networks worth considering, especially given computational costs [151]. Currently, SNNs are not yet as accurate in comparison to traditional neural networks, although they are more bio-inspired (they have characteristics that are more similar to biological neurons) [152]. Zhou et al. [153] applied SNNs to the classification of malignant melanomas and benign melanocytic nevi (skin lesions with disorder classification). Average accuracy was achieved and the computational cost was lower than that of CNNs. Toğaçar et al. [154] also found that skin cancer could automatically be detected using SNNs. In other studies, drawing on SNN-based image segmentation, it was shown that an MRI image scan segmentation could diagnose dementia [155,156].

Results showed that this approach made it possible to predict changes in the brain image (taking into account mild cognitive impairment) two years before they occurred with an accuracy of 90 percent. Thus, SNNs provide the opportunity to evaluate change dynamics, in this case of the brain, and similar results were obtained by Turkson et al. [157]. This is significant because in the case of neurodegenerative diseases, diagnosing changes in the brain is often very difficult without effective MRI analysis, although neurodegenerative diseases can also be diagnosed less invasively using EEG. Here, however, SNNs can also be used to analyze the results obtained [158]. Another example of the use of SNN networks is the detection of epileptic seizures [159], classification of motor tasks [160], and automatic sleep staging [161].

*5.1.8 Generative Adversarial Networks*

Generative Adversarial Networks (GANs) are network architectures that consist of two core components: the generator and the discriminator. The generator shoulders the responsibility of creating data that faithfully emulates specific data (artificial data identical to real data) so as to cheat the discriminator. It initiates the process with an input of random noise, meticulously refining it through multiple layers of neural network architecture. Each layer integrated within the generator network fulfills a distinct role, harnessing techniques such as convolutional or fully connected layers. These layers operate cohesively to progressively metamorphose the initial noise input into an output that becomes increasingly indistinguishable from the target data. A discriminator is designed to distinguish artificial data (produced by a generator) from real data based on small nuances. Thus, the core concept of this solution is to train two networks that compete with each other. As a consequence, they are expected to produce more authentic data [162,163]. This ongoing adversarial interaction propels both networks towards continuous enhancement, resulting in increasingly sophisticated capabilities in data generation and validation. In the context of medical imaging, the role of the generator holds paramount significance [164].

*5.1.9 Graph Neural Networks*

If the data format is approached differently, as in non-Euclidean space in the form of graphs, it can be understood in terms of vertices (i.e. objects). Then, the concept of Graph Neural Networks (GNNs) can be applied [165]. All relations in this type of neural network are expressed as those between nodes and edges of the graph. These networks are designed to handle graph data that form a critical aspect in medical fields, especially when the intricate relationships and connections between data points are essential for accurate diagnosis and health condition analysis. This principle of operation is useful in medical imaging, especially in neuroimaging and molecular imaging, where understanding complex relationships is crucial. GNNs function through message transmission or aggregation, enabling nodes to gather information from their neighbors. This learning process, grounded in the local structure and features of the graph, is iterative and allows the network to understand broader contexts within the graph, which is essential in comprehending complex biological structures and functions. Central to GNN architecture are graph convolutional layers. These layers are designed to handle signals over the nodes and edges of a graph, enabling the network to capture the relationships and features inherent in the graph’s structure [166]. Through these layers, each node aggregates information from its neighbors, thus capturing local graph structures. This is a critical aspect of GNNs as it allows for a nuanced understanding of how each part of the graph relates to its surroundings. Following graph convolutional layers come pooling layers. These layers aggregate information from groups of nodes into a single node, effectively reducing the graph's complexity. This dimensionality reduction is important as it helps significant features in the graph to be emphasized while also reducing computational load. Pooling layers, therefore, play a significant role in summarizing information contained in the graph, making it manageable and more interpretable. The architecture also includes fully connected layers. After the data is processed through the earlier layers, it passes through these layers where the high-level, aggregated features from the graph are synthesized. These layers are essential in integrating the learned features from the graph, transforming them into a format suitable for the final output, be it for prediction, classification, or any other specific task. The output layer is the last component in the GNN architecture. This layer is responsible for generating the final output of the network, which can vary depending on the specific application [167,168].

*5.1.10 Sinusoidal Representation Networks*

Sinusoidal Representation Networks also called SIRENs are a new type of activation function, that is to say, a periodic activation function that is applied to an implicit neural representation instead of the commonly used rectified linear unit (ReLU) and sigmoid activation functions. The SIREN's basic operation principles are based on multiplying the previous result (input data that passes through different layers) by the weight matrix and then submitting the activation function. The results (output from the last network layer) are an approximation of the desired function. The type of activation function used has a positive impact on the calculation optimization process (computational cost) due to smooth derivatives. Sitzmann et al. [169] proposed this solution to the image processing task issue. Considering the complexity of medical images, this is an interesting direction for further research.

*5.1.11 Transformers*

One further type of neural network that has recently come into focus in the field of medicine concerns Transformers. They learn rules based on learning the context and tracking the relations between the data [138]. Originally, these were networks used for Natural Language Processing (NLP). Their effectiveness in these tasks resulted in the development of transformers such as the Detection Transformer (DETR) for tasks related to vision analysis [170], the Swin-Transformer [53], the Vision Transformer (ViT) [171], and the Data-Efficient Image Transformer (DeiT) [172]. The DETR is dedicated to object detection which also includes manual analytical processes, and it uses CNN to learn 2D representations of the input data (images). In turn, the ViT converts input to a series of fixed-size non-overlapping patches and treats them as a token. Each of them encodes the spatial position of each part of the image to provide spatial information, while the spatial information of the pixels is lost during tokenization. However, ViTs require large training datasets. On the other hand, DeiTs also provide high accuracy in the case of small training datasets, while Swin-Transformers allow the cost of calculations to be reduced. They process an image divided into overlapping areas showing tokens at multiple scales with a hierarchical structure using a shifted window (local self-attention). The transformer principle of operation is based on the self-attention mechanism. This enables the network to decide on the importance of different parts of the input data for future prediction (i.e. weight). This may be beneficial for the evaluation of the relationships between different regions in medical images. For example, the majority of AI-based MRI analysis is performed employing CNNs. However, this introduces the lack of long-range dependency as a limitation. The application of the Swin transformer for this task was therefore proposed [53]. Additionally, Usam et al. [173] proposed a ViT-based analysis of chest radiography and pediatric pneumonia, while Gheflati and Rivaz [174] suggested its use for the classification of breast ultrasound images. The disadvantage of transformers in the area of medical image segmentation concerns the lack of large datasets. To overcome this disadvantage, modifications to their architecture have been introduced, such as those proposed by Roy et al. [175], and other possible solutions include the addition of further types of neural networks such as CNNs [176,177,178].

*5.2.12 Quantum Neural Networks*

Recently, some work has also been devoted to the development of quantum neural networks (QNNs) that are based on the idea of quantum mechanics [179,180]. These may have a huge potential to speed up calculations and reduce the computational costs associated with them. This approach can be developed in two ways related to the segmentation of medical images. The first is the use of quantum circuits to train classical neural networks, and the second is the design and training of quantum networks, as proposed by Mathur et al [134]. Indeed, Shahwar et al. [181] showed the potential of QNNs in the classification of Alzheimer's detection, and Ullah et al. [182] proposed a quantum version of the Fully Convolutional Neural Network (FCNN) as applied to a challenge that concerned the classification of ischemic heart disease. This allowed for a prediction accuracy of over 80 percent. However, the approach based on quantum neural networks requires further improvement. When it comes to interventional practice, QNNs have the potential for stenosis detection in X-ray coronary angiography [183], and they can be also applied to selecting medicines for patients with high accuracy [184,185].

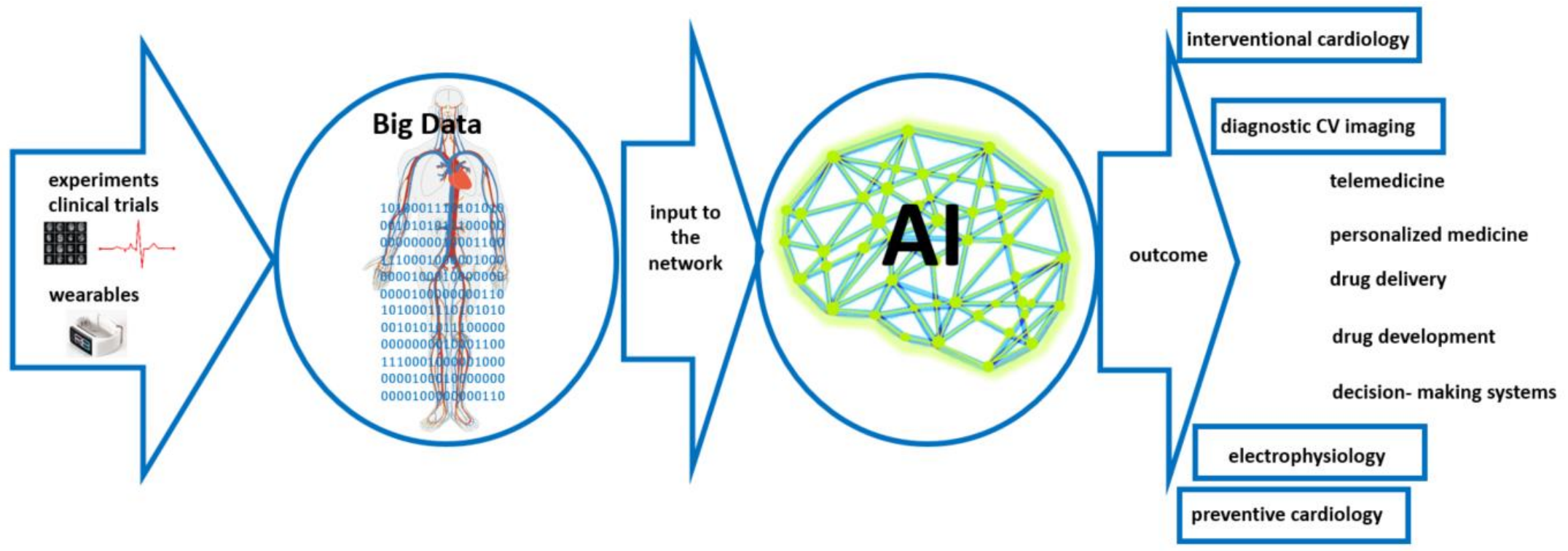

**Figure 3**. Conceptual scheme of the application of AI in cardiology.

*5.2. Artificial Intelligence in cardiology*

Mathematical and physical models can bridge the gap between experimental measurements and theoretical sciences. Since the first breakthroughs in classical physics, with Galileo Galilei and Isaac Newton being the most prominent figures, mathematical modeling has revealed its unparalleled explanatory power in describing phenomena observed in the surrounding world. A further and even more significant consequence of simulating reality is the emerging predictive power, which enables us to foresee potential experimental outcomes and, based on this knowledge, optimally adjust experimental designs. Computer-assisted medicine in general and cardiac modeling, in particular, is by no means an exception from the successful application of continuous advancements in bioelectricity and biomagnetism [186]. Along with enhancements in measuring techniques of the ECG and a constant increase of computational resources, these advances have provoked the development of many different heart models that can support an automatic and accurate diagnosis of the heart, beat by beat. Knowledge of the anatomical heart structure is an important part of the evaluation of cardiac functionality. Thus, cardiac images are one of the significant techniques applied in the assessment of patient health. The image segmentation procedure is usually performed manually, with an expert sitting in front of a monitor moving a pointer, and not only does this require time and resources to accomplish, but it is also subject to error depending on the experience of the expert. Indeed, this procedure is time-consuming, inefficient, very often error-prone, and highly user-dependent [138]. Therefore, the development of an efficient, automatic segmentation procedure is of great importance [65]. However, certain limitations result in the fact that the automatic segmentation of cardiac images is still an open and difficult task. For example, in the case of 2D echocardiographic images low signal-to-noise ratio, speckles, and low-quality images form some of the difficulties in determining the contour of the ventricles. Moreover, significant variability in the shape of heart structures makes it difficult to develop universal automated algorithms. The basic concept of AI application in cardiology is presented in **Figure 3**.

*5.2.1 Application of the YOLO Algorithm*

In advancing the diagnosis of cardiovascular diseases (CVDs), the YOLOv3 algorithm was also developed for the precise segmentation of the left ventricle (LV) in echocardiography. This method leverages YOLOv3's powerful feature extraction capabilities to accurately locate key areas of the LV, including the apex and bottom, facilitating the acquisition of detailed LV subimages. Employing the Markov random field (MRF) model for initial identification and processing, the method then applies sophisticated techniques including nonlinear least-squares curve fitting for exact LV endocardium segmentation. YOLOv3's role is pivotal in ensuring the accuracy and efficiency of this process, highlighting its significance in the early detection and analysis of CVD [187]. On the other hand, in the realm of cardiac health monitoring and medical image processing, the Lion-Based Butterfly Optimization model with Improved YOLO-v4 was introduced as described by Alamelu & Thilagamani, [188]. When applied in the prediction of heart disease based on echocardiography, it resulted that a refined version of the segmentation algorithm significantly improves (with an average of 99% accuracy) the analysis of echocardiographic images, offering more accurate and thorough insights into cardiac health, thus marking a substantial advancement in cardiac diagnostics technology. Lee et al. [189] applied the YOLO-v5 algorithm to cardiac detection. Based on cardiovascular CT images from Soonchunhyang University Hospital in Korea, the critical role of data preprocessing in deep learning, especially when dealing with limited medical datasets, was presented. The study highlights the advanced capabilities of YOLO-v5, including its efficiency in processing and analyzing complex medical images, in particular in the field of cardiology. The approach presented significantly contributes to the precision and speed of cardiac disease detection, underscoring the impact of deep learning techniques in improving medical diagnostics. Moreover, by combining the capabilities of YOLOv7 with a U-Net Convolutional Neural Network the precise segmentation of left heart structures from echocardiographic images was subsequently developed [190]. This approach efficiently delineates complex anatomical structures, including the left atrium, endocardium, and epicardium. Thus, the integration of YOLOv7 with U-Net significantly improves the accuracy and efficiency of the segmentation process, proving to be a valuable asset in cardiac diagnosis, Animesh Tandontics, and clinical practice.

*5.2.2 Genetic Algorithms*

Genetic Algorithms (GAs) can also be used to improve diagnosis as well as selection of targeted therapy in the field of cardiology. Reddy et al. [191] applied GAs to the diagnostics of early-stage heart disease, which has crucial implications in the selection of further therapy methods. For example, GAs allowed for the optimization of classification rules. As a consequence, the level of accuracy increased and the computational cost was reduced (due to simplification of the selection process). GAs can also be applied to the determination of personalized parameters of the cardiomyocyte electrophysiology model [192]. The Cauchy mutation was applied. In most cases, GAs were used to limit the number of parameters that are then used as input to another AI-based algorithm, such as, for example, a Support Vector Machine (SVM) [193,194].

*5.2.3 Artificial Neural Networks*

In cardiology, the ability to detect conditions accurately and at an early stage is of paramount importance, and the application of ANNs for the analysis of medical images is an important development in this area. Considering the high global prevalence of cardiovascular diseases, the application of ANNs in cardiac imaging may substantially improve diagnostic techniques [195]. ANNs provide an efficient computational tool to detect structural abnormalities in heart tissues [196]. They also play a vital role in assessing cardiac function, evaluating important metrics such as ejection fraction, and analysis of blood flow patterns, essential for diagnosing heart failure or valvular heart disease. Their ability to analyze historical and current medical images aids in predicting the progression of cardiac diseases. This could positively impact patient outcomes, meeting an essential requirement in contemporary healthcare. Based on ANNs, Salte et al. [197] proposed automating the measurement of global longitudinal strain (GLS), a vital metric for assessing left ventricular function in cardiology. Echocardiographic cine-loops were analyzed and the approach developed demonstrates superior accuracy and efficiency compared to conventional speckle-tracking software. A further study by Nithyakalyani et al. [198] also shows ANN potential in the CVD diagnostic process.

*5.2.4 Convolutional Neural Networks*

The role of CNNs in addressing cardiovascular diseases is substantial and influential. For example, Roy et al. [175] applied CNNs to cardiac image segmentation to diagnose Coronary Artery Disease (CAD). CNNs were used to analyze 2D X-ray images, significantly enhancing image segmentation accuracy and setting new standards in medical image analysis. Similarly, as in Gao et al. [199], Galea et al. [200] proposed combining U-Net and DeepLabV3+ CNN architectures for the segmentation of cardiac images from smaller datasets. Tandon et al. [201] applied CNNs in cardiology with a specific focus on cardiovascular imaging for patients with Repaired Tetralogy of Fallot (RTOF). A CNN originally designed for ventricular contouring was retrained and adapted to the complexities of RTOF. This enabled an increase in algorithm accuracy. In turn, Stough et al. [202] developed a fully automatic method for segmenting heart substructures in 2D echocardiography images using CNNs that was validated against a robust dataset, and Sander & Išgum [203] focused on enhancing the segmentation of cardiac structures in cardiac MRI. This method integrates automatic segmentation with an assessment of segmentation uncertainty to identify potential local failures. The measures of predictive uncertainty were calculated and trained by another CNN to detect local segmentation errors for potential expert correction. This approach combining automatic segmentation with manual correction of detected errors could significantly reduce the time required for expert segmentation. Masutani et al. [204] considered the lengthy acquisition times and reduced spatial detail in cardiac MRI. Here, CNNs were applied for deep learning super-resolution. It turned out that CNNs considerably outperformed traditional image upscaling methods, recovering high-frequency spatial details and providing accurate left ventricular volumes. Then, Liu et al. [125]

focused on creating interpretable deep-learning models for cardiac MRI segmentation, particularly of the left ventricle with the use of CNNs. A deep CNN was also applied to classify coronary computed tomography angiography (CCTA) scans using the Coronary Artery Disease Reporting and Data System (CAD-RADS) categories [196]. Indeed, one of the advantages of this approach is the reduction of the analysis time compared to manual readings, demonstrating its efficiency and accuracy in automating the classification process for coronary artery disease. For example, O'Brien et al. [205] proposed automated detection of ischemic scars in the left ventricle from routine CTA imaging. They aimed to integrate seamlessly with existing clinical workflows, using only anatomical information. The importance of this study lies in its ability to detect scar tissue, crucial for accurate diagnosis and intervention planning in cardiac patients. The CNN exhibited high accuracy in detecting scar slices, performing better than manual readings and showing the potential of this method in enhancing cardiac imaging and diagnostics at minimal additional costs. Similarly, Candemir et al. [206] employed a deep learning algorithm using a 3-dimensional CNN to detect and localize coronary artery atherosclerosis in CCTA scans. By extracting coronary arteries and utilizing a saliency map to visualize potential atherosclerosis locations, this algorithm demonstrated high accuracy and effectiveness. Its significant negative predictive value indicates its usefulness in assisting clinicians, particularly in acute chest pain cases, to rule out coronary atherosclerosis.

*5.2.5 Recurrent Neural Networks*

In the case of medical data in the form of echocardiography, Recurrent Neural Networks (RNNs) are integral in processing and analyzing heart imaging data. Their ability to effectively capture information from sequential image data significantly contributes to the accuracy of cardiac analysis. This is essential in understanding the continuous nature of the heart and the subtle structural changes that occur, ensuring a comprehensive and detailed examination of cardiac health [207]. Similarly, in cardiac MRI segmentation, RNNs have also shown promising performance [208]. MRI and CT are pivotal in cardiology, utilizing magnetic fields and radio-frequency waves for MRI, and X-ray computed tomography for CT. These technologies offer non-invasive and comprehensive insights into the heart's anatomy and function. RNNs excel in handling the sequential and temporal aspects of both MRI and CT data, crucial for monitoring dynamic changes in cardiac tissues over time. The proficiency of RNNs in analyzing MRI and CT images significantly contributes to cardiology, enhancing the diagnosis and ongoing management of heart conditions [209]. In turn, Wahlang et al. [210] combined RNNs and LSTM successfully in the segmentation and classification of 2D echo images, 3D Doppler images, and video graphic images. Wang & Zhang 211 also considered the segmentation of the left ventricle wall in four-chamber view cardiac sequential images. Recognizing the complexity of the four-chamber structure and wall motion diversity, their study introduces a dense RNN algorithm for accurate LV wall segmentation in four-chamber view MRI time sequences. Employing two RNNs, the first for providing detailed information for the initial image and the second for generating the segmentation result, this approach increases the accuracy of the first frame image and the overall effectiveness of information flow between LSTM cells. Another RRN application in the field of cardiology was presented by Muraki et al. [212]. Here, simple RNNs, LSTM, and GRU were used to detect acute myocardial infarction (AMI) in echocardiography. The study demonstrates the potential of RNNs in improving the accuracy and reliability of AMI detection, potentially leading to better treatment outcomes for cardiac patients. Another cardiology-connected RNN application field was developed by Fischer et al. [213]. A deep learning algorithm, combining RNN with LSTM was adopted to detect coronary artery calcium (CAC) from coronary computed tomography angiography (CCTA) data. The automatic detection and labeling of heart and coronary artery centerlines was considered. The RNN-LSTM algorithm focused on determining CAC presence, achieving high diagnostic accuracy, sensitivity, and specificity. This approach may offer an efficient, automated approach to identifying calcified plaques, crucial in coronary artery disease management. Lyu et al. [214] put forward a recurrent generative adversarial network model for cine cardiac magnetic resonance imaging. This model utilizes bidirectional convolutional LSTM and multi-scale convolutions, adept at managing long-range temporal features and capturing both local and global features, thereby enhancing the network's performance. The method showcases significant improvements in cine cardiac MRI image quality and an ability to generate missing intermediate frames, thus improving the temporal resolution of cine cardiac MRI sequences. A similar approach can be found in the work of Ammar et al. [215] concerning a lightweight U-Net variant with a 2D convolutional LSTM layer, effectively handling the sequential nature of cine MRI 3D spatial volumes. This provides good segmentation accuracy (over 0.9 DICE coefficient) in the case of the left and right ventricle cavities (LVC and RVC, respectively), and the left ventricle myocardium (LVM). Additionally, the method shows strong correlation coefficients and limits of agreement for clinical indices when compared to their ground truth counterparts, highlighting its potential effectiveness and efficiency in cardiac cine MRI analysis.

*5.2.6 Spiking Neural Networks*

Calculations related to the analysis of cardiac data are very time-consuming and involve a great deal of computing resources. One alternative that can potentially reduce computational cost could be Spiking Neural Networks (SNNs). They may also be advantageous in wearable and implantable devices for their energy efficiency and real-time processing capabilities. This makes them ideal for continuous cardiac monitoring, as they require less frequent recharging or battery replacement, a significant benefit for devices like cardiac monitors and pacemakers. For example, Rana and Kim [216] modify the synaptic weights such as to be binary. This operation provides a reduction in computational complexity and power consumption. This is crucial, especially in the context of wearable monitors where continuous monitoring is key but the constraints of power and computational resources are limiting factors. Their binarized SNN model emerges as not just a viable, but a highly efficient alternative for ECG classification, setting a new standard in continuous cardiac health monitoring technologies. SNNs also provide a diagnostic tool worth considering, as found in Shekhawat et al. [217]. They propose a Binarized Spiking Neural Network (BSNN) optimized with a Momentum Search Algorithm (MSA) for fetal arrhythmia detection. Moreover, the incorporation of a Momentum Seach Algorithm (MSA) significantly boosts the classification accuracy of the BSNN, underlining the adaptability and effectiveness of SNNs in handling intricate biomedical signals. Another study in the field of arrhythmia detection introduces a Memristive Spike-Based Computing in Memory (MSB-CIM) system using a Memristive Spike-Based Computing Engine With Adaptive Neuron (MSPAN). Then, a multi-layer deep integrative spiking neural network (DiSNN) in edge computing environments was developed by Jiang et al. [118]. This system efficiently manages ECG classification tasks, greatly reducing computational complexity without compromising accuracy. The implementation of a memristor-based CIM architecture enhances the system's compactness and power efficiency, making it an ideal solution for edge device applications where size and energy consumption are critical considerations. Furthermore, Banerjee et al. [218] optimized SNNs for ECG classification in wearable and implantable devices such as smartwatches and pacemakers. Their approach in designing both reservoir-based and feed-forward SNNs, and integrating a new peak-based spike encoder, has led to significant enhancements in network efficiency. A feed-forward SNN, when combined with a peak-based encoder, stands out as the most efficient model. It strikes an optimal balance between computational cost and energy consumption, making it a highly suitable system for wearable edge devices. Yan et al. [63] proposed training the SNN model on diverse patient data and then adeptly applied it to classify ECG patterns from new, untrained individuals. This approach addresses the balance between low power consumption and high accuracy effectively, making it a highly suitable choice for continuous, real-time heart monitoring in everyday wearable technology. Similarly, Kovács & Samiee [219] introduced a hybrid neural network architecture that merges the strengths of Variable Projections (VP) with the capabilities of SNNs. They focused on creating a robust and efficient architecture where the VP layer, in tandem with spiking layers, encodes the input space into a compact and interpretable latent feature space, namely a Variable Projection Spiking Neural Network (VPSNN). Thus, the ability to classify complex ECG patterns accurately using this hybrid model could lead to significant advancements in the early detection and treatment of cardiac conditions. An interesting solution in ECG classification has also been presented by Feng et al. [220]. Their approach involves building a structure analogous to a deep ANN, transferring the trained parameters to this new structure, and utilizing leaky integrate-and-fire (LiF) neurons for activation. This method not only matches but in some cases, exceeds the accuracy of the original ANN model. This may lead to more efficient, accurate, and reliable systems for continuous cardiac health monitoring, potentially revolutionizing the way heart diseases are detected and monitored.

*5.2.7 Generative Adversarial Networks*

Generative Adversarial Networks (GANs) seem to be promising computational tools to elevate patient care and improve clinical outcomes, in particular in the field of cardiology. First, the most important GAN application field is CVD diagnosis [221]. Retinal fundus images were used as input to the network. This approach led to the analysis of microstructural alterations within retinal blood vessels to pinpoint pivotal risk factors associated with CVD, such as Hypertensive Retinopathy (HR) and Cholesterol-Embolization Syndrome (CES). Moreover, the incorporation of a retrained ImageNet model for customized image classification further bolstered predictive accuracy, offering invaluable solutions for healthcare professionals in terms of treatment strategies and risk mitigation. Furthermore, Chen et al. [209] demonstrated the potential of GAN in automating precise segmentation of the left atrium (LA) and atrial scars in late gadolinium-enhanced cardiac magnetic resonance (LGE CMR) images. The quantification of atrial scars, distinguished by substantial volume disparities, necessitated a departure from traditional two-phase segmentation methods. To surmount this hurdle, JAS-GAN was introduced, an inter-cascade generative adversarial network, to autonomously and accurately segment unbalanced atrial targets within LGE CMR images. Thus, an adaptive attention cascade and adversarial regularization culminate in simultaneous and precise segmentation of both LA and atrial scars. This solution provides some insight into clinical applications in the treatment of patients with atrial fibrillation (AF), underscoring the indispensable role of GANs in the realm of medical imaging tasks. The transformative potential of GANs to enhance dynamic CT angiography derived from CT perfusion data has been shown in studies by Wu et al. [222]. Vessel-GAN, characterized as an explainable generative adversarial network, allows for standalone coronary CT angiography. This contributes to augmenting patient safety by mitigating ionizing radiation and contrasting agent usage. Additionally, automated atherosclerosis screening from coronary CT angiography (CCTA) by harnessing the capabilities of Generative Adversarial Networks (GANs) was developed by Laidi et al. [223]. GANs help to address the conundrum of limited positive images within the test dataset. The generation of new images reminiscent of the original dataset enables bolstered sensitivity and Positive Predictive Value (PPV). While the proposed model exhibited promising outcomes, a marginal dip in accuracy and specificity after the incorporation of generated images underscored the need for further fine-tuning. Zhang et al. [224] concentrated on the precise segmentation of ventricles within MRI scans. Their work recognized the difficulties posed by unclear contrast, blurred boundaries, and noise inherent in these images. They compared the effectiveness of two segmentation methods, U-Net and Transfer Learning and Multi-Scale Discriminant Generative Adversarial Network (TLMDB GAN), ultimately finding that TLMDB GAN outperformed U-Net in terms of segmentation accuracy. Pushing development further, Decourt & Duong [225] addressed the essential task of left ventricle segmentation in pediatric MRI scans. They introduced DT-GAN, a GAN approach that uses semi-supervised semantic segmentation to reduce the reliance on large annotated datasets. Their innovative GAN loss function and methodology enhanced segmentation accuracy, particularly for boundary pixels, showing promise for automated left ventricle segmentation in cardiac MRI scans. Diller et al. [226] explored the potential application of Progressive Generative Adversarial Networks (PG-GAN) to generate synthetic cardiac MRI images for congenital heart disease research. They successfully trained a PG-GAN using the MRI frames of a prospective nationwide study, creating synthetic images suitable for training segmentation networks such as U-Net. This approach enables both data privacy concerns to be addressed and yields segmentation results comparable to those achieved with direct patient data, showcasing the potential of PG-GANs in generating realistic cardiac MRI images for rare cardiac conditions.

*5.2.8 Graph Neural Networks*

In the field of cardiology, GNNs have been effectively employed in several key areas. They have been used in the classification of polar maps in cardiac perfusion imaging, a critical technique for assessing heart muscle activity and blood flow. Another significant application of GNNs in cardiology is the estimation of left ventricular ejection fraction in echocardiography. This measurement is vital for evaluating heart health, specifically in assessing the volume of blood the left ventricle pumps out with each contraction [227]. This allows for more accurate analyses through an understanding of the intricate graph structures of the heart's imagery. GNNs are also being utilized in analyzing CT/MRI scans. This approach can also be used to interpret the relationships and structures within the scan, providing detailed insights into various conditions and helping in diagnosis and treatment planning [125]. A further application of GNNs in cardiology is connected with cardiac perfusion imaging. This task covers the classification of the polar maps which is necessary for the evaluation of heart muscle activity and blood flow. These maps also play an important role in echocardiography, particularly in the estimation of the left ventricular ejection fraction, an important indicator of heart function. GNNs have also been applied in predicting ventricular arrhythmia and segmenting cardiac fibrosis based on MRI data [228], a two-stage deep learning network for segmenting the left ventricle myocardium and fibrosis in Late Gadolinium Enhanced (LGE) CMR images, achieving high dice scores and surpassing previous methods. This approach may provide potential improvements in ventricular arrhythmia treatment and SCD risk assessment. Ping Lu et al. [228] proposed Spatio-Temporal Graph Convolutional Networks (ST-GCNs) to diagnose cardiac conditions, namely by understanding and quantifying left ventricular (LV) motion in cardiac MR cine images. Another GNN application field is that of the automated anatomical labeling of coronary arteries, which addresses the variability of human anatomy [229]. This approach was based on a Conditional Partial-Residual Graph Convolutional Network (CPR-GCN). This is a combination of 3D CNNs AND LSTM. Fan Huang et al. [230] focused on predicting Coronary Artery Disease (CAD) from CT scans using vascular biomarkers derived from fundus photographs through a GNN. This method showed that specific retinal vascular biomarkers, such as arterial width and fractal dimensions, were significantly associated with adverse CAD-RADS scores. Simultaneously, Gao et al. [231] tackled the automation of coronary artery analysis using Coronary Computed Tomography Angiography (CCTA). This crucial analysis assists clinicians in diagnosing and evaluating CAD. Deep learning models are used for centerline extraction and lumen segmentation of coronary arteries. Ome of the components, a CNNTracker, traced the coronary artery centerline, while a Vascular Graph Convolutional Network (VGCN) achieved precise lumen segmentation. This method included an iterative refinement process alternating between the CNNTracker and GCN. It resulted in providing a high level of accuracy in CCTA data analysis from patients, particularly in key arteries such as the right coronary artery (RCA), left coronary artery (LCA), and X-ray coronary angiography (XCA). In another study, a GNN-based method for comorbidity-aware chest radiograph screening was developed [232]. It allows the screening of cardio, thoracic, and pulmonary conditions to be enhanced, and in this way significantly improves screening performance over standard ensemble techniques. Another interesting study introduced the Nonlinear Regression Convolutional Encoder-Decoder (NRCED), a framework designed to map multivariate inputs to multivariate output [233]. This framework was specifically applied to the reconstruction of 12-lead surface ECG from intracardiac electrograms (EGMs) and vice versa. The study analyzed the features learned by the model, utilizing them to create a diagnostic tool for identifying atypical and diseased heartbeats. A high Receiver Operating Characteristic (ROC) curve is produced with an associated area under the curve (AUC) value of 0.98, indicating excellent discrimination between the two classes. This approach may have a significant potential for improving cardiac patient monitoring and diagnostics, ultimately enhancing healthcare outcomes.

*5.2.9 Transformers*

The application of Transformer networks allows for a deeper understanding of cardiac function, which aids in refining diagnostic methods and improving treatment strategies. For example, Jungiewicz et al. focused on stenosis detection in coronary arteries, comparing different variants of the Inception Network with the ViT [234]. They analyzed small fragments from coronary angiography videos, highlighting the role of dataset configuration in model performance. A key innovation in their approach is the use of Sharpness-Aware Minimization (SAM) alongside Vision Transformers (VTs), which enhances the accuracy and reliability of stenosis detection. They also employed explainable AI techniques to understand the differences in classification performance between the models. Their findings indicate that while convolutional neural networks generally outperform transformer-based architectures, the gap narrows significantly with the addition of SAM to VTs. In some measures, the SAM-VT model even surpasses other models. It turned out that ViT can effectively be applied to diagnose coronary angiography. Zhang et al. [235] present a Topological Transformer Network (TTN) for automated coronary artery branch labeling in cardiac CT angiography (CCTA). The TTN, inspired by the success of transformers in sequence data analysis, treats vessel branch labeling as a sequence labeling learning problem. It introduces a unique topological encoding to represent spatial positions of vessel segments within the arterial tree, enhancing classification accuracy. The network also includes a segment-depth loss function to address the class imbalance between primary and secondary branches. The effectiveness of a TTN is demonstrated in CCTA scans, where it achieves unprecedented results, outperforming existing methods in overall branch labeling and side branch identification. TTNs mark a departure from traditional methods, representing the first transformer-based vessel branch labeling method in the field. The integration of this method into computer-aided diagnosis systems can enhance the generation of cardiovascular disease diagnosis reports, thereby improving patient outcomes in cardiac care. Additionally, Minqi Liao et al. [236] proposed a novel approach for Left Ventricle (LV) segmentation in echocardiography using pure transformer models. They proposed two models: one combining the Swin transformer with K-Net and another utilizing Segformer, evaluated on the EchoNet-Dynamic dataset. These models excel in segmenting challenging cardiac regions, such as the valve area, and separating the left ventricle from the left atrium, particularly in difficult samples. This work fully utilizes the capabilities of the transformer architecture for LV segmentation, moving beyond traditional methods and showcasing the potential of transformers in clinical applications. While the study currently focuses on static frames without including automated LVEF calculation, the researchers plan to extend these models to echocardiographic videos in future work. This represents a significant advancement in medical imaging, particularly in cardiac echocardiography, demonstrating the powerful applications of TNNs in healthcare technology. Going further, Ahn et al. [237] introduced the Co-Attention Spatial Transformer Network (CA-STN) for unsupervised motion tracking in 3D echocardiography.

This approach significantly enhances the detection and analysis of myocardial ischemia and infarction by tracking wall-motion abnormalities in the left ventricle. The core innovation is the integration of a co-attention mechanism within the Spatial Transformer Network (STN), which improves feature extraction between frames for smoother motion fields and enhanced interpretability in noisy 3D echocardiography images. Additionally, a novel temporal regularization term guides the motion of the left ventricle, producing smooth and realistic cardiac displacement paths. The CA-STN outperforms traditional methods that rely on heavy regularization functions, marking a new standard in cardiac motion tracking. Strain analysis using the Co-Attention STNs aligns with matched SPECT perfusion maps, illustrating the clinical utility of 3D echocardiography for localizing and quantifying myocardial strain following ischemic injury. This study contributes a novel tool for cardiac imaging and opens new possibilities for early detection and interventions in myocardial injuries. In another advancement, Lhuqita Fazry et al. [238] developed a groundbreaking approach using hierarchical vision transformers to estimate cardiac ejection fraction from echocardiogram videos. Addressing the variability in ejection fraction assessment among different observers, this method does not require prior segmentation of the left ventricle, making it a more efficient process. The team's evaluations on the EchoNet-Dynamic dataset show enhanced accuracy and efficiency compared to state-of-the-art methods, demonstrating the potential of TNNs in revolutionizing cardiac function assessment. The public availability of their source code fosters further innovation in the field. Yang Ning et al. [239] proposed Efficient Multi-Scale Vision Transformers (EMVTs) for coronary artery calcium (CAC) segmentation (CAC-EMVT). This approach addresses the segmentation of CAC, which often has fuzzy boundaries and inconsistent appearances. The CAC-EMVT effectively models both short and long-range dependencies using a combination of local and global branches. Its three distinct modules, Key Factor Sampling (KFS), Non-Local Sparse Context Fusion (NSCF), and Non-Local Multi-Scale Context Aggregation (NMCA), enhance segmentation accuracy. Tested on CT scans from CVD patients, the CAC-EMVT shows significant improvements over existing methods in accuracy and reliability, representing a significant step forward in the detection and analysis of coronary artery calcium. On the other hand, Han et al. [240] developed a method for detecting coronary artery stenosis in X-ray angiography (XRA) images. Their hybrid architecture, integrating transformer neural networks with convolutional neural networks (CNNs), captures the spatio-temporal nuances of XRA sequences. The Proposal-Shifted Spatio-Temporal Tokenization (PSSTT) within the transformer module tokenizes spatio-temporal elements of XRA sequences processed through the Transformer-based feature aggregation (TFA) network. Erwan et al. [241] introduced a new method for segmenting cardiac infarction in delayed-enhancement MRI, tackling the challenge of differentiating between healthy and infarcted myocardial tissues. Their approach, aimed at enhancing the quantitative evaluation of myocardial infarction using Late Gadolinium Enhancement cardiac MRI (LGE-MRI), employs a dual-approach methodology. Initially, a dedicated 2D U-Net generates a probability map of the healthy myocardium, which guides the accurate localization of infarcted areas. Then, a U-Net transformer network refines this segmentation by combining the probability map with the original image. An adapted loss function addresses the limitations of U-Net in segmenting infarcted regions, significantly improving accuracy. Similarly, Ding, et al. [242] developed a transformative approach for segmenting and classifying myocardial fibrosis in DE-MRI scans. Addressing the complex process of categorizing fibrotic tissue, their self-supervised myocardial histology segmentation algorithm employs a Siamese system for multi-scale representation. The integration of an end-to-end method using a transformer model for detecting myocardial fibrosis tissue is a key feature. This model combines a Pre-LN Transformer with a Multi-Scale Transformer (MST) backbone and a joint regression cost to accurately determine distances between forecast blocks and labels. The method significantly improves performance metrics, establishing its effectiveness and reliability in segmenting and classifying myocardial fibrosis. In turn, Upendra et al. [243] proposed a hybrid architecture combining ViT for deformable image registration of 3D cine cardiac MRI images. This approach consistently estimates cardiac motion by capturing the optical flow representation between consecutive 3D volumes from a 4D cine cardiac MRI dataset. Experiments on the Automated Cardiac Diagnosis Challenge (ACDC) dataset demonstrate superior results in deformable image registration compared to traditional methods. This advancement showcases the potential of vision transformers in enhancing the accuracy and reliability of cardiac function assessment, representing a major stride in cardiac imaging technology.

a) b)

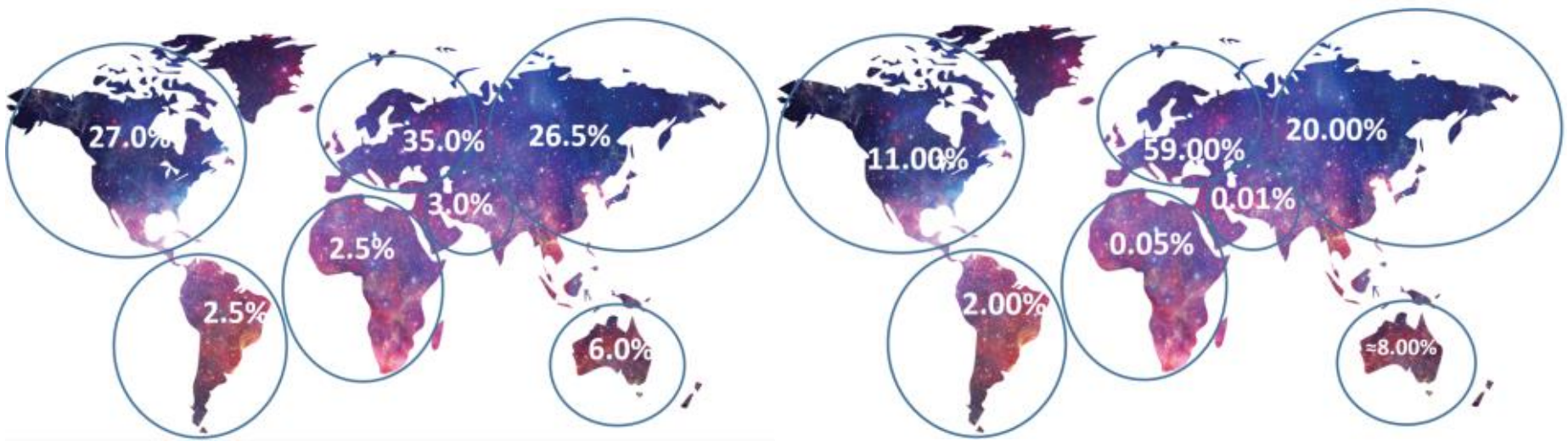

c)

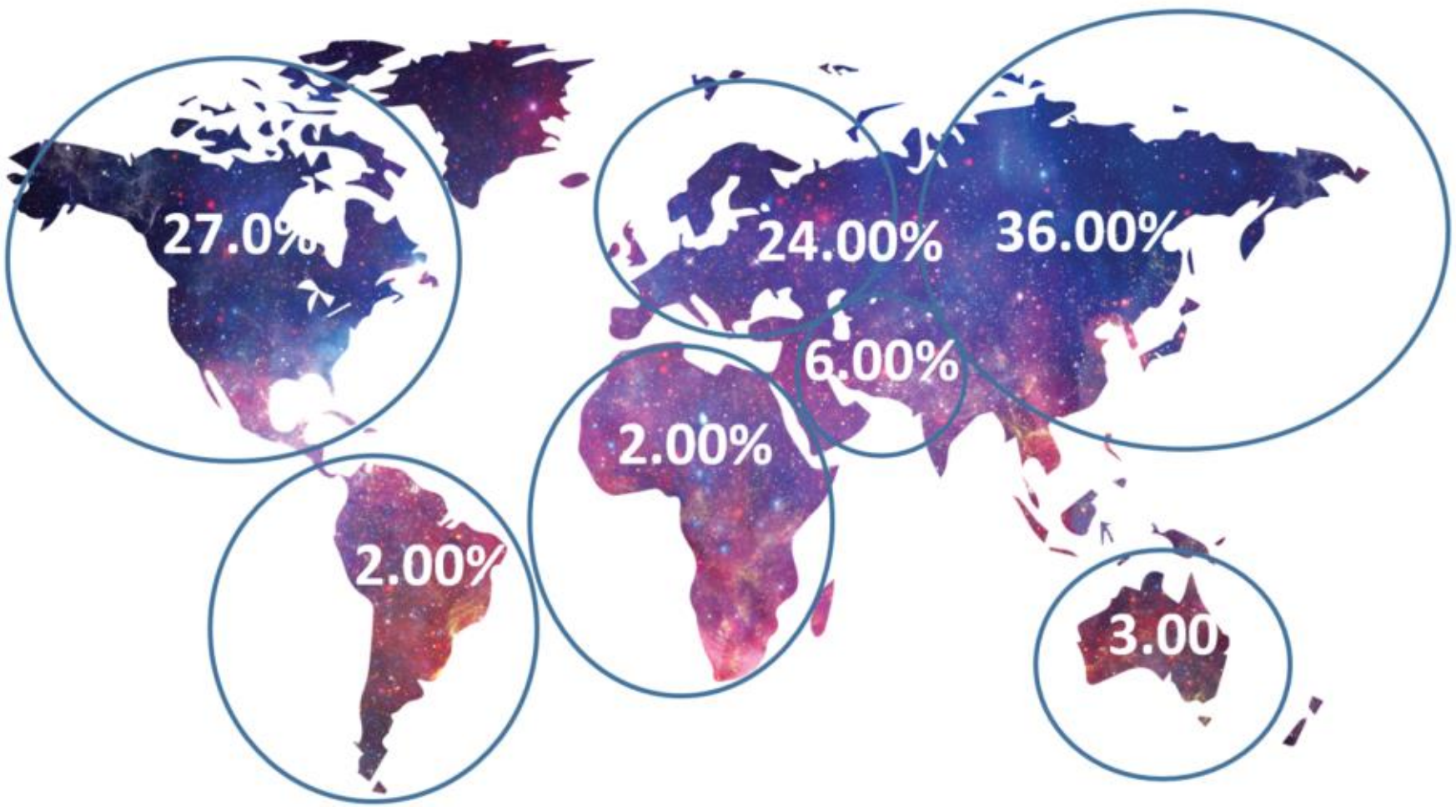

**Figure 4.** Distribution of papers in the field of medicine based on origin a) in the field of DT-technologies, b) in the field of XR-based technologies, and c) in AI-based algorithms in image segmentation (WoS database).

### *5.3 Evaluation Metrics in Medical Image Segmentation*

Artificial intelligence has the chance to become a high-precision tool in medicine. However, there are certain technical risks (TERs) connected with the application of AI in clinical and educational practice, including algorithm performance, legal regulation, and safety. For example, it is known that small, even imperceptible changes in the training dataset can drastically change the results of predictions, which in medicine can have very serious consequences and influence learning. The key to the evaluation of AI adaptability is to use an appropriate metric to assess the correctness and accuracy of different kinds of forecasts including clinical prognoses and for this to be understood by users [244]. For example, overfitting between training and testing datasets will reduce the accuracy of the algorithm. Other crucial factors that influence the qualitative efficiency of the AI-based algorithm's data set include data availability issues. However, even if developers do not have sufficient quantity and quality of data, cross-validation can be applied [245]. This procedure helps avoid overfitting by the selection of a subset. Thus, the choice of a proper evaluation metric depends on the specific task type. The binary classifier Dice coefficient (also called Sørensen–Dice index) and Index of Union (IoU) are most commonly used in medical image segmentation metrics. However, in the field of cardiology accuracy is of particular concern (see **Table 3**).

**Table 2**. Top list of used AI models in cardiology, including interventional cardiology.

| AI/ML model | Application Fields (in general) | Application Fields (in cardiology) | References |
|---|---|---|---|
| *ANNs* | classification, pattern recognition, image recognition, natural language processing (NLP), speech recognition, recommendation systems, prediction, cybersecurity, object manipulation, path planning, sensor fusion | prediction of atrial fibrillation, acute myocardial infarctions, and dilated cardiomyopathy detection of the structural abnormalities in heart tissues | 246<br>196 |
| *RNNs* | ordinal or temporal problems (language translation, speech recognition, NLP image captioning), time series prediction, music generation, video analysis, patient monitoring, disease progression prediction | segmentation of the heart and subtle structural changes<br>cardiac MRI segmentation | 207<br>208 |
| *LSTMs* | ordinal or temporal problems (language translation, speech recognition, NLP, image captioning), time series prediction, music generation, video analysis, patient monitoring, disease progression prediction | segmentation and classification of 2D echo images<br>segmentation and classification of 3D Doppler images<br>segmentation and classification of video graphics images<br>detection of the AMI in echocardiography | 210<br>210<br>210<br>212 |
| *CNNs* | pattern recognition, segmentation/classification, object detection, semantic segmentation, facial recognition, medical imaging, gesture recognition, video analysis | cardiac image segmentation to diagnose CAD<br>cardiac image segmentation to diagnose tetralogy of Fallot<br>localization of the coronary artery atherosclerosis | 175, 247<br>248<br>206 |
| | | detection of cardiovascular abnormalities | 249 |
| | | detection of arrhythmia | 182, 250 251, 252, 253, 254, 255, 256, 257, 258 |
| | | detection of coronary artery disease | 218, 259 |
| | | prediction of the survival status of heart failure patients | 260 |
| | | prediction of cardiovascular disease | 261 |
| | | LV dysfunction screening | 262 |
| | | prediction of premature ventricular contraction detection | 263,264 |
| *Transformers* | NLP, speech processing, computer vision, graph-based tasks, electronic health records, building conversational AI systems and chatbots | coronary artery labeling | 235, 234 |
| | | prediction of incident heart failure | 265 |
| | | arrhythmia classification | 266, 267, 268, 211 |
| | | cardiac abnormality detection | 269 |
| | | segmentation of MRI in case of cardiac infarction | 270 |
| | | classification of aortic stenosis severity | 271,272, 272 |
| | | LV segmentation | 273, 267 239 |
| | | heart murmur detection | 274 |
| | | myocardial fibrosis segmentation | 267 |
| | | ECG classification | 275 |
| *SNNs* | pattern recognition, cognitive robotics, SNN hardware, brain-machine interfaces, neuromorphic computing | ECG classification | 276, 167, 277 |
| | | detection of arrhythmia | 219, 278, 279 |
| | | extraction of ECG features | 280 |
| *GANs* | image-to-image translation, image synthesis and generation, data generation for training, data augmentation, creating realistic scenes | CVD diagnosis | 221 |
| | | segmentation of the LA and atrial scars in LGE CMR images | 209 |
| | | segmentation of ventricles based on MRI scans | 224 |
| | | left ventricle segmentation in pediatric MRI scans | 225 |
| | | generation of synthetic cardiac MRI images for congenital heart disease research | 226 |
| *GNNs* | graph/node classification, link prediction, graph generation, social/biological network analysis, fraud detection, recommendation systems | classification of polar maps in cardiac perfusion imaging | 281,282 |
| | | analysis of CT/MRI scans | 125 |
| | | prediction of ventricular arrhythmia | 282 |
| | | segmentation of cardiac fibrosis | |
| | | diagnosis of cardiac condition: LV motion in cardiac MR cine images | 282 |
| | | automated anatomical labeling of coronary arteries | 283 |
| | | prediction of CAD | 229 |
| | | automation of coronary artery analysis using CCTA | 230 |
| | | screening of cardio, thoracic, and pulmonary conditions in chest radiograph | 231 |
| | | | 232 |
| *QNNs* | optimization of hardware operations, user interfaces | classification of ischemic heart disease | 182 |
| *GA* | optimization techniques, risk prediction, gene therapies, medicine development | classification of heart disease | 191 |

## 6. Data Security Issues Connected with the Metaverse and Artificial Intelligence

In the era of rapid development of artificial intelligence, the metaverse, and digital twins, a natural question that arises concerns the field of data security and this is particularly crucial in medicine. In view of this, new approaches such as AI Trust, Risk, and Security Management (AI TRiSM) have been developed [284]. This framework enables an AI-based system to be evaluated according to certain criteria such as compliance, fairness, reliability, and preserving data privacy. Data security in an AI-based system is quite complicated. It includes security of systems design, model testing, applications, regulatory compliance, infrastructure, auditing, and an ethics review. Thus, medical data can become subject to attacks, both passive and active. The way medical systems such as implantable medical devices and internet wearable devices are implemented makes them more vulnerable to attacks than other systems. It is possible that as many as half of all attacks are carried out in this sector [285]. Given that human health or even life is at stake, these systems must be specially protected. Data must be subject to authentication, availability, integrity, non-repudiation, and confidentiality. To minimize the leakage of sensitive information about the patient, a process of anonymization is intended to prevent such information about the patient from being read again based on his or her medical record including patient identification. To this end, the most common method involves the use of pseudonymization techniques such as replacing direct identifiers with pseudonymous codes. When it comes to the security of medical information systems many solutions involve blockchain technologies [286] involving robust encryption and authentication methods [287]. Also, the idea of storage and distribution of sensitive information among the number of cloud nodes combined with encryption has been introduced [288], based on quantum deep neural networks. It has turned out that this approach provides a better detection rate than other commonly applied methods. Moreover, all medical systems that process and store sensitive personal information must be developed and used with the compliance of the European General Data Protection Regulation (GDPR) and the American California Consumer Privacy Act (CCPA) [289]. However, differences between regulations in Europe and the USA have effectively hampered the exchange of sensitive patient information without appropriate institutional safeguards [290]. In the case of patient privacy, XR-based systems also provide good solutions [291]. Furthermore, AI can also be applied to tracking attacks and their location.

**Table 3**. Comparison of AI models applied to cardiology, including interventional cardiology.

| Network Type | Evaluation metrics | Input | Output | XR connection | DT Contention | Reference |
|---|---|---|---|---|---|---|
| ANN | accuracy 94.32% | ECG recordings | binary classification of normal and ventricular ectopic beats | No | No | 292 |
| ANN | ROI 89.00% | Echocardiography | automatic measurement of left ventricular strain | No | No | 197 |
| ANN | accuracy 91.00% | Electronic Health Records | classification and prediction of cardiovascular diseases | No | No | 195 |
| RNN-LSTM | accuracy 80.00%<br>F1 Score 84.00% | 3D Doppler images | heart abnormalities classification | No | No | 210 |
| RNN-LSTM | accuracy 97.00%<br>F1 Score 97.00% | 2D echo images | heart abnormalities classification | No | No | 210 |
| RNN-LSTM | 1:accu-raqcy 85.10%<br>2:accuracy 83.20% | echocardiography images | automated classification of acute myocardial infarction:<br>1:classification of the left ventricular long-axis view<br>2:classification of short-axis view (papillary muscle level) | No | No | 212 |
| RNN-LSTM | accuracy 93.10% | coronary computed tomography angiography | diagnostic of the coronary artery calcium | No | No | 213 |
| RNN-LSTM | accuracy 90.67% | ECG recordings | prediction of the arrhythmia | No | No | 250 |
| RNN | IoU factor 92.13% | MRI cardiac images | estimation of the cardiac state: sequential four-chamber view left ventricle wall segmentation | No | No | 211 |
| CNN | accuracy 95.92% | ECG recordings | binary classification of normal and ventricular ectopic beats | No | No | 292 |
| CNN | IoU factor 61.75% | MRI cardiac images | estimation of the cardiac state: sequential four-chamber view left ventricle wall segmentation | No | No | 211 |
| CNN | accuracy 94.00%<br>F1 Score 95.00% | 2D echo images | heart abnormalities classification | No | No | 210 |
| CNN | accuracy 98.00%<br>F1 Score 98.00% | 3D Doppler images | heart abnormalities classification | No | No | 210 |
| CNN | accuracy 92.00% | ECG recordings | ECG classification | No | No | 63 |
| CNN | accuracy 88.00% | Electronic Health Records | heart disease prediction | No | No | 221 |
| CNN | accuracy 98.82% | ECG recordings | prediction of heart failure and arrhythmia | No | No | 254 |
| CNN | accuracy 95.13% | Electronic Health Records | prediction of the survival status of heart failure patients | No | No | 260 |
| CNN | accuracy 99.60% | ECG recordings | estimation of the fetal heart rate | No | No | 293 |
| CNN | accuracy 99.10% | heart audio recordings | heart disease classification | No | No | 294 |
| CNN | accuracy 97.00% | heart sound signals | classification of heart Murmur | No | No | 295 |
| CNN | accuracy 98.95% | heart sound signals | classification of heart sound signals | No | No | 296 |
| CNN | ROC curve (AUC) 0.834 | heart sound signals | prediction of obstructive coronary artery disease | No | No | 297 |
| CNN | accuracy 85.25% | MRI image scans | chronic disease prediction | No | No | 298 |
| CNN | accuracy 99.10% | heart sound signals | diagnosis of cardiovascular disease | No | No | 299 |
| CNN-LSTM | accuracy 99.52%<br>Dice coefficient 0.989<br>ROC curve (AUC) 0.999 | ECG recordings | prediction of congestive heart failure | No | No | 300 |
| CNN-LSTM | accuracy 96.66% | Heart disease Cleveland UCI dataset | prediction of the heart disease | No | No | 301 |
| CNN-LSTM | accuracy 99.00% | ECG recordings | prediction of the heart failure | No | No | 258 |
| SNN | ROC curve (AUC) 0.99 | ECG recordings | ECG classification | No | No | 278 |
| SNN | accuracy 97.16% | ECG recordings | binary classification of normal and ventricular ectopic beats | No | No | 292 |
| SNN | accuracy 93.60% | ECG recordings | ECG classification | No | No | 118 |
| SNN | accuracy 85.00% | ECG recordings | ECG classification | No | No | 216 |
| SNN | accuracy 84.41% | ECG recordings | ECG classification | No | No | 276 |
| SNN | accuracy 91.00% | ECG recordings | ECG classification | No | No | 63 |
| GNN | Dice coefficient 0.82 | cardiac MRI images | prediction diverticular arrhythmia | No | No | 282 |
| GNN | ROC curve (AUC) 0.739 | CT image scan | prediction of coronary artery disease | No | No | 230 |
| GNN | AUC 0.821 | chest radiographs | screening of cardio, thoracic, and pulmonary conditions | No | No | 232 |
| GNN | ROC area 0.98 | 12-lead ECG recordings | remote monitoring of surface electrocardiograms | No | No | 233 |
| GAN | accuracy 99.08%<br>Dice coefficient 0.987 | CT image scan | cardiac fat segmentation | No | No | 44 |
| GAN | accuracy 98.00% | ECG recordings | ECG classification | No | No | 302 |
| GAN | accuracy 95.40% | ECG recordings | ECG classification | No | No | 303 |
| GAN | accuracy 68.07% | CTG signal dataset | fetal heart rate signal classification | No | No | 304 |
| GAN | Dice coefficient 0.880 | MRI image scans | segmentation of the left ventricle | No | No | 225 |
| Transformers | accuracy 96.51% | Cleveland dataset | prediction of cardiovascular diseases | No | No | 305 |
| Transformers | accuracy 98.70% | heart sound signals - Mel-spectrogram, Bispectral analysis, and Phonocardiogram | heart sound classification | No | No | 149 |
| Transformers | Dice coefficient 0.861 | 12-lead ECG recordings | arrhythmia classification | No | No | 122 |
| Transformers | Dice coefficient 0.0004 | ECG recordings | arrhythmia classification | No | No | 306 |
| Trans-formers | Dice coefficient 0.980 | ECG recordings | arrhythmia classification | No | No | 307 |
| Trans-formers | Dice coefficient 0.911 | ECG recordings | classification of ECG recordings | No | No | 280 |
| GA | - | laboratory data, patient medical history, ECG, physical examinations, and echocardiogram (Z-Alizadeh Sani dataset) | determination of the parameters to prediction of the coronary artery disease (next SVM-based classifier was applied) | No | No | 194 |
| QNN | accuracy 84.60% | electronic health records | classification of ischemic cardiopathy | No | No | 182 |
| QNN | accuracy 91.80% Dice coefficient 0.918 | X-ray coronary angiography | stenosis detection | No | No | 183 |

## 7. Ethical issues connected with the Metaverse and Artificial Intelligence

Within medicine as a whole, the Metaverse is conceived both as a general space where the behavior of medical practitioners in their use of technologies such as XR, VR, AR, and MR could be subject to the same ethical issues and open to the same ethical threats as found in other virtual landscapes such as online gaming 308 and as a specific space where such technologies are used to develop and deploy treatment practices for specific diseases such as those covered by cardiology. In the former, ethical issues such as moral equivalence arise. For example, Grinbaum [309] asks whether behavior in the metaverse should be judged according to real-world values, or whether there are aspects of virtual behavior that are new or overlapping: how should actors who assume different personae be treated when those personae act badly in different ways? Then, Radovanović & Tomić [310] point out that actors entering the metaverse may come with a set of religious beliefs that may color their judgments and actions, such as how they set up a virtual church. These beliefs will likely also develop as the actor's life in the virtual world unfolds, and very general codes of conduct have been proposed in this case. Indeed, early attempts have been made to establish ethical codes of conduct as practical guidelines for humans operating avatars in the metaverse. The practical code of ethics proposed by Heider [311] is aimed at humans operating in the metaverse through their avatar(s). The code has seven points: show respect, tell the truth, do no harm, show concern, work for good, demonstrate tolerance, and respect privacy. However, only the first two of these seven specifically deal with behavior that is qualitatively different from that found in the physical world: the others have general applicability. They highlight two important aspects of difference: a human may evoke or create more than one avatar with different appearances, and those may assume different roles. In the latter conception of the metaverse as a specific space, physicians will use particular metaverse technologies to deal with certain diseases and conditions, where there is a contrast between a tool or suite of tools designed to do a job [312] and a general environment in which virtual agents may live [313,314]. However, in both conceptions of the metaverse, similar ethical issues are found: human characteristics inform behavior, attitudes, and usage. The application of Digital Twins (DTs) has become central in medical practice overall. Applications are found in areas ranging from clinical trials to treatment interventions, medical education, and scenario modeling [315]. In all of these cases, the two basic notions of the metaverse are core: an environment and a set of tools along with issues of representation. Indeed, Braun [316], raises the difficult issue of how a person is to be represented by their DT in terms of accuracy (in the various types of representation) and control (who will have the authority to control the DT and how?). Safeguarding issues also arise with regard to children [316] and, by extension, vulnerable groups of people.

Clinical DTs in cardiology provide the opportunity for the generation of abundant quantities of data. Armeni [315] points out that DTs comprise virtual modeling of qualitatively different types of real-world objects and phenomena ranging from people to devices, environmental features, and institutions (such as clinics) connected by means of data streams used not only during treatment interventions but also in clinical trial design and medical education. All these are relevant to cardiology. For example, the production of data during treatment (whether rapid (emergency interventions) or slow (cardiac monitoring) might be used not only to inform patient choice but also to make financial decisions and even set insurance premiums: heart issues comprise an important part of any medical insurance form. However, the very notion of a clearly defined dataset comes into question when the data journey is considered: data travels from one location to another and may be adapted along the way [317]. Thus, the identification and description of cardiac-related data may prove ethically contentious. Moreover, issues such as security and privacy, data characteristics (selection, collection, categorization, and use), ownership, and rights of use and access all come into play and all comprise risk points. Unless they are carefully defined, the ability to make insurance or even liability claims (in case of potential clinical malpractice) may be limited and problematic. Overall, DT models will become progressively more detailed and accurate, such as those of the human heart described by Viola et al. [312] and those reviewed by Coorey et al. [318,319]. However, ethical dimensions are currently lacking in such models. These increasingly important dimensions include not only ownership and control of data but also the human side of the influence of the DT on the human and the rights of stakeholders. A truly effective DT will need four parts: the physical, the virtual, the data connection, and the ethical. Indeed, neglect of the ethical side could be said to be the biggest threat to the development of DT technology, especially in the vital area of cardiology. In this perspective, there is a clear need for DTs to be explainable through a framework of Explainable AI (XAI) and trustworthy through one of Trustworthy AI (TAI) [319]. If development in these areas is lacking, progress on the technical side will be held up due to a series of multiple ethical objections. These factors can be accounted for at some level, whether as integrated into specific systems, in local policies, or in national and international regulations [320].

## 8. Discussion and Conclusions

Extended Reality provides a tool for 3D representation of the structures of the heart [45]. Although HDMs offer great opportunities in clinical cardiology, they are not without drawbacks. Some users complain of health problems after long-term use of HDMs, including dizziness, nausea, and even blurred vision (symptoms accompanying motion sickness) [321,322], although this cybersickness may not be experienced by some users [323,324]. Also, the application of AI-based algorithms to six degrees of freedom motion support in a VR simulator may help alleviate cybersickness [325]. Hardware improvements such as frame rates and headset tracking have made it possible to partially counteract symptoms, but further development of HDMs is needed. On the other hand, Daling et al. [326] showed that XR-based training is not necessarily better, but at least as good as traditional methods. Another significant limitation of HDMs is the size of the field of view, much smaller than the field of view of a human. In turn, the implementation of XR-based solutions in clinical practice is also limited due to their high cost, in particular in lower-middle-income countries (LMICs) [327]. In **Figure 2** and **Figure 3**, the percentage share of latest published research by territory (2020-2024) in AI, XR, and DTs is presented. It can be seen that LMICs have a low participation level. However, the spread of XR as a tool seems to be inevitable, especially in medical practice.

Since HDTs can generate specific data, they can also predict the outcome of a surgical procedure, disease progression, or the performance of an implanted device. HDTs combined with AI and XR-based technology also unlock the potential for sustainable development of the healthcare ecosystem. In cardiology, a personalized, computational model of the heart is crucial in better understanding patient-specific pathophysiology and supporting clinical decision-making processes, but the development of the heart DT requires the fitting of various types of parameters, including cardiac electromechanics and cardiovascular hemodynamics parameters [328]. Indeed, the implementation of the heart digital twin is a complex process, which has not yet been fully accomplished [329]. Thus, further research activity should be concentrated on issues of applying an electrical cardiac modeling approach in combination with artificial intelligence-based algorithms to build a digital twin of the heart for different clinical applications, ranging from those used by the general practitioner to the highly specialized electrophysiologist.

Furthermore, AI-based algorithms have been successfully used in recent medical imaging, in particular in the field of cardiology. Consequently, a list of AI models used in cardiology, including interventional cardiology, according to application field is presented in **Table 2**, and it has been shown that CNNs and Transformers are the most frequently used solutions in the field of cardiology, while GA is commonly used to optimize the parameter space. **Table 3** provides a summary of the types of neural networks used in cardiology, taking into account their accuracy and application area, and it also includes information on the relationship between the neural network and XR and DT-based technology. It became evident that only a few studies combine these fields, and then regarding certain concepts and perspectives [34].

Cardiovascular diseases, encompassing various ailments like coronary artery disease, heart failure, hypertensive heart disease, and many others, represent a significant health burden globally. When the application of AI in cardiology is considered, the use of ANNs in diagnosing these conditions could lead to more effective management and treatment strategies, potentially reducing the global impact of CVD [198]. However, ANN application in the field of medical image processing requires converting two-dimensional photos to one-dimensional vectors. This increases the number of parameters and increases the cost of calculation. However, RNNs have proven to be well-suited to managing the sequential and temporal characteristics inherent in MRI and CT data, a capability that is essential for accurately tracking the dynamic alterations in cardiac tissues due to the possibility of effective capturing of long-range non-linear dependencies, such as modeling the risk trajectory of heart failure [330]. However, the limitation of RNNs is connected with vanishing or exploding gradients, although CNNs offer users even better performance results than humans can achieve [144]. Their characteristics and capabilities contribute to the fact that they are the most common choice for medical image segmentation procedures. On the other hand, SIRENs are more resistant to the vanishing gradient problem. However, more complex architecture may be needed to achieve the same accuracy as networks with a different activation function.

In the context of cardiology, the fully connected layers of CNNs are responsible for synthesizing information to perform critical analytical tasks. These include classifying different cardiac conditions, detecting anomalies such as irregularities in heart size or shape, and making predictive assessments based on the comprehensive analysis of cardiac structure and function. CNNs are particularly good at handling complex datasets from various imaging modalities in cardiology, including MRI, CT scans, and ultrasound [177]. Their capacity to automatically learn and identify feature representations, without manual intervention, marks a significant advancement in the field of cardiology. The strength of CNNs lies in their ability to handle high-dimensional data and to effectively capture the spatial structures within medical images in cardiology. This leads to more precise and comprehensive analyses of cardiac health. As CNN technology evolves, it is poised to play an increasingly crucial role in the accurate and efficient examination of medical images, leading the way to more sophisticated and personalized healthcare solutions [331]. However, important limitations include the gridded input format. In the case of sparse or partial input data, their use is difficult and does not provide high prediction accuracy. High segmentation accuracy is associated with high computational costs. Learning spatial correlations requires the use of many network layers with a large number of neurons, which also involves significant computational resources. Additionally, object surfaces require post-processing heuristics to achieve smoothness. Nor do CNNs take into account spatial relations in images which is important in the case of medical ones [332]. To overcome this limitation, Capsule Networks (CNs) were introduced [333]. Their output is in the form of vectors that enable some spatial relations to be saved. The disadvantage of this approach is the lack of verification on a large dataset. Another disadvantage of CNNs is the high computation cost. To reduce this cost, Visual Transformers (VTs) were introduced [138]. However, they may require larger datasets than traditional CNNs to achieve the same accuracy. Transformers are also able to capture non-local patterns and relationships between pixels. As a consequence, they can take into account complex spatial dependencies in medical images. The results of analyses conducted with transformers may be complex and hard to interpret. Other disadvantages of transformers may be connected with difficulties in capturing hierarchical spatial hierarchies.

On the other hand, GANs have shown exceptional proficiency in handling complex and varied cardiac datasets. They generate highly realistic images, aiding training and research, particularly where access to real patient data is limited. GANs are instrumental in enlarging existing datasets and creating diverse and extensive data for training more accurate and robust diagnostic models. Besides image generation, GANs are adept at image-to-image translation tasks, a significant feature in medical imaging [334]. They can transform MRI images into CT scans, offering different perspectives of the same anatomical structure without needing multiple imaging modalities. This is particularly beneficial in scenarios where certain imaging equipment might be unavailable. As GAN technology continues to advance, its role in medical imaging, especially in cardiology, is set to expand. The intricate architecture of GANs, with their dual-network system and adversarial training approach, allows for the generation and refinement of data. This opens up new avenues for research, training, and diagnostic methods, leading to more personalized and advanced healthcare solutions. The ongoing development of GANs promises to further our understanding and interpretation of complex medical data, heralding a new era in healthcare technology [335]. The main disadvantages of GANs are the complex training needed that does not necessarily lead to hoped-for results, a tendency to overfit, and high computational costs. Additionally, GANs are difficult to interpret, which is of key importance in medicine, especially in cardiology.

GNNs provide a powerful tool for understanding and interpreting complex data structures, such as those found in medical image processing [124]. One of the key strengths of GNNs is their adaptability to varying input sizes and structures, an essential feature in medical imaging where patient data can greatly differ. The architecture of GNNs is tailored to process and interpret graph-structured data, making it a powerful tool in areas such as medical image processing where data often forms complex networks. This specialized structure of GNNs sets them apart in their ability to handle data that is inherently interconnected, such as neurological networks or molecular structures. It is also worth stressing that GNNs were created for tasks that cannot be effectively solved by other types of networks based on input data in Euclidean space. Moreover, GNNs are difficult to interpret. On the other, computational cost is also a crucial parameter. Here, QNNs may provide some insight, while the GA can effectively help in the optimization of the input parameters to neural networks. Another possibility is based on the application of SNNs, which are more computationally efficient (connected to the high level of computational speed and real-time performance). As a consequence, SNNs consume less energy, which translates into better use of hardware resources. However, their learning algorithms require improvement (in terms of accuracy gains), in comparison for example to the accuracies achieved by the application of CNNs [151]. In the case of SNNs, the requirement of increasingly powerful hardware is also of high importance.

While there have been and continue to be great technical advances in the specific technologies of HDTs, ethical concerns are not generally systematically connected to the same. Rather, the ethical discussion tends to take place in parallel with the technical, whereas a more robust model would integrate the two. Ethical issues ranging from control and ownership of data to social values embedded in technical decisions and human behavior in the Metaverse need to be addressed at every step along the way. If this does not happen, progress may be delayed and even blocked in some cases by ethical disputes, thus holding back valuable DT applications in cardiology and other areas of medicine.

The development of HDTs in pre-clinical imaging gives numerous benefits, including improved outcomes, shorter development timelines, and lower costs. The application of HDTs will be increasingly popular in the future of healthcare service and has a huge potential to become central in the mainstream of medicine. However, it requires the development of both models and algorithms for the analysis of medical data. On the other hand, in cardiology, the interpretation of ECGs currently relies on experts and requires training and clinical expertise and is thus subject to considerable inter and intra-clinician variability. Additionally, the diagnostic value of the standard 12-lead ECG is limited by the difficulty of linking the ECG data directly to cardiac anatomy and also due to the prevalence of technical errors such as electrode positioning. Therefore, the combination of AI, XR, and HDT technology in cardiology with the potential of avoiding technical errors can serve as a universal methodology to predict health status and improve outcome quality.

**Supplementary Materials:** Not applicable.

**Author Contributions:** "Conceptualization, A.P., K.P., M.P. and P.D.; methodology, A.P., and Z.R.; formal analysis, A.P., K.P., P.D., and Z.R.; investigation, A.P., K.P., P.D., M.P., and Z.R.; resources, A.P., K.P., P.D., and Z.R.; data curation, A.P., and Z.P.; writing—original draft preparation, A.P., Z.P., and M.P.; writing—review and editing, A.P., K.P., P.D., M.P., and Z.R.; visualization, A.P.; supervision, A.P. All authors have read and agreed to the published version of the manuscript."

**Funding:** "This research received no external funding".

**Data Availability Statement:** Not applicable.

**Acknowledgments:** This study was supported and financed by the National Centre for Research and Development under Grant Lider No. LIDER/17/0064/L-11/19/NCBR/2020. The research was also partially supported by the National Centre for Research and Development (research grant Infostrateg I/0042/2021-00).

**Conflicts of Interest:** The authors declare no conflicts of interest.